\documentclass[a4paper,11pt,preprintnumbers]{article}
\pdfoutput=1 
\usepackage{jcappub} 
\usepackage{booktabs}
\usepackage[T1]{fontenc} 
\usepackage{color,subfigure}
\usepackage{graphicx, multirow,soul,url,amsmath,amsfonts,amssymb,mathrsfs,amsfonts}
\usepackage[all]{hypcap}
\usepackage{url}
\usepackage{bbm}
\usepackage{cancel}
\usepackage{braket}
\usepackage{subfigure}
\usepackage[normalem]{ulem}
\usepackage{slashed}
\usepackage[dvipsnames]{xcolor}
\usepackage{multicol, blindtext}
\usepackage[section]{placeins}
\usepackage[version=3]{mhchem}
\usepackage{caption}
\usepackage{lipsum}
\usepackage{hyperref}
\usepackage{float}
\usepackage{mattens}
\usepackage{mathtools}
\usepackage{footnote}
\usepackage{amsmath,amssymb,amsthm,amsxtra,overpic,bbm,bm,epsfig}
\usepackage{color,subfigure}
\usepackage[splitrule]{footmisc}
\usepackage{bbold}
\usepackage{amsmath}
\usepackage{amsbsy}
\usepackage[export]{adjustbox}
\usepackage{units}
\usepackage{commath}

\newcommand{\be}{\begin{equation}}
\newcommand{\ee}{\end{equation}}
\newcommand{\bea}{\begin{eqnarray}}
\newcommand{\eea}{\end{eqnarray}}
\newcommand{\ba}{\begin{aligned}}
\newcommand{\ea}{\end{aligned}}

\newcommand{\equaref}[1]{Eq.~(\ref{#1})}

\newcommand{\figsref}[2]{Figs.~\ref{#1}~and~\ref{#2}}

\newcommand{\MPBH}{{M}}
\newcommand{\MPBHini}{{M}_{\rm ini}}
\newcommand{\Tform}{\tilde{T}_{\rm form}}
\newcommand{\Tevap}{\tilde{T}_{\rm evap}}
\newcommand{\alphaform}{N_e^{\rm form}}

\newcommand{\MN}{M_N}

\newcommand{\YB}{{Y}_{\rm B}}
\newcommand{\YBobs}{{Y}_{\rm B}^{\rm obs}}

\newcommand{\NPBH}{\mathcal{N}_{\rm PBH}}

\newcommand{\rhotot}{\rho_{\rm tot}}
\newcommand{\rhorad}{\rho_{\rm rad}}

\newcommand{\Mpl}{{M}_{\rm Pl}}

\newcommand{\Tsphal}{T_{\rm sphal}}
\newcommand{\Tewpt}{T_{\rm EWPT}}

\newcommand{\rcrit}{r_{\rm core}}
\newcommand{\rdec}{r_{\rm dec}}

\newcommand{\Tcore}{T_{\rm core}}
\newcommand{\Tplasma}{\tilde{T}}
\newcommand{\TBH}{T_{\rm BH}}
\newcommand{\rsphal}{r_{\rm sphal}}
\newcommand{\rcore}{r_{\rm core}}

\newcommand{\X}{$X$ }

\newcommand{\INFN}{INFN - Sezione di Napoli, Complesso Univ. Monte S. Angelo, I-80126 Napoli, Italy}
\newcommand{\UNINA}{Dipartimento di Fisica ``Ettore Pancini'', Università degli studi di Napoli ``Federico II'', Complesso Univ. Monte S. Angelo, I-80126 Napoli, Italy}
\newcommand{\IPPP}{Institute for Particle Physics Phenomenology, Durham University, South Road, DH1 3LE, Durham, United Kingdom}
\newcommand{\KCL}{King’s College London, Strand, London, WC2R 2LS, United Kingdom}

\title{Primordial Black Hole Hot Spots and Out-of-Equilibrium Dynamics}
\author[a,b]{Jacob Gunn}
\author[c]{, Lucien Heurtier}
\author[d]{, Yuber F.~Perez-Gonzalez}
\author[d]{and Jessica Turner}

\affiliation[a]{\UNINA}
\affiliation[b]{\INFN}
\affiliation[c]{\KCL}
\affiliation[d]{\IPPP}

\preprintnumber{\newline KCL-PH-TH-2024-48, IPPP/24/58}

\emailAdd{jacobwilliam.gunn@unina.it}
\emailAdd{lucien.heurtier@kcl.ac.uk}
\emailAdd{yuber.f.perez-gonzalez@durham.ac.uk}
\emailAdd{jessica.turner@durham.ac.uk}

\abstract{
When light primordial black holes (PBHs) evaporate in the early Universe, they locally reheat the surrounding plasma, creating hot spots with temperatures that can be significantly higher than the average plasma temperature. 
In this work, we provide a general framework for calculating the probability that a particle interacting with the Standard Model can escape the hot spot. More specifically, we consider how these hot spots influence the generation of the baryon asymmetry of the Universe (BAU) in leptogenesis scenarios, as well as the production of dark matter (DM). For leptogenesis, we find that PBH-produced right-handed neutrinos can contribute to the BAU even if the temperature of the Universe is below the electroweak phase transition temperature, since sphaleron processes may still be active within the hot spot.
For DM, particles emitted by PBHs may thermalise with the heated plasma within the hot spot, effectively preventing them from contributing to the observed relic abundance. Our work highlights the importance of including hot spots in the interplay of PBHs and early Universe observables.}


\begin{document}
\maketitle

\def\thefootnote{\arabic{footnote}}
\setcounter{footnote}{0}


\section{Introduction}
Two of the most important unresolved questions in particle physics and cosmology are the nature of Dark Matter (DM) and the origin of the Baryon Asymmetry of the Universe (BAU), both of which have received considerable attention. 
According to standard lore, the production mechanisms for both DM and the BAU rely on out-of-equilibrium conditions that likely emerged as the early radiation-dominated Universe cooled after a (brief) period of reheating following inflation.
One of the most well-investigated mechanisms for DM and BAU production involves processes once in thermal equilibrium with the primordial plasma falling out of equilibrium as the expansion rate exceeded the interaction rates. 

Nevertheless, little is known about the Universe before the Big Bang Nucleosynthesis (BBN) era.
The Universe could have experienced a matter-dominated phase or there could have been further sources of entropy that diluted any preexisting DM or baryon populations. 
Additionally, separate non-thermal mechanisms for producing DM and the BAU could have been present.
A notable example of a non-standard cosmological history that features all these different possibilities simultaneously is the presence of an evaporating primordial black hole (PBH) population in the early Universe~\cite{Carr:1974nx,Carr:1975qj,Carr:1976zz,Khlopov:2008qy}.
The potential presence of PBHs and their impact on the evolution of the Universe has garnered significant interest, particularly following the discovery of gravitational waves (GW) and the confirmation of the existence of black holes~\cite{LIGOScientific:2016aoc}.
Unlike stellar black holes, PBHs can be extremely light (heavier than the Planck mass), and therefore rapidly radiate away their mass in a process known as Hawking radiation \cite{Hawking:1974rv,Hawking:1975vcx,Hawking:1976ra}. Since black holes are democratic radiators, they would produce any beyond the Standard Model states related to DM or the BAU, and several works have explored the production of DM~\cite{Lennon:2017tqq,Morrison:2018xla,Hooper:2019gtx,Auffinger:2020afu,Gondolo:2020uqv,Bernal:2020bjf,Bernal:2020ili,Bernal:2020kse,Baldes:2020nuv,Masina:2020xhk,Masina:2021zpu,Sandick:2021gew,Bernal:2021bbv,Bernal:2021yyb,Cheek:2021odj,Cheek:2021cfe,Barman:2021ost,Bernal:2022oha,Cheek:2022mmy,Chen:2023lnj,Chen:2023tzd,Kim:2023ixo,Barman:2024iht,Barman:2024slw,Haque:2023awl,RiajulHaque:2023cqe} or Dark Radiation~\cite{Hooper:2019gtx,Lunardini:2019zob,Masina:2020xhk,Masina:2021zpu,Hooper:2020evu,Arbey:2021ysg,Cheek:2022dbx,Eby:2024mhd}, baryogenesis either independently~\cite{Barrow:1990he,Majumdar:1995yr,Upadhyay:1999vk,Dolgov:2000ht,Bugaev:2001xr,Baumann:2007yr,Hooper:2020otu} or through leptogenesis~\cite{Fujita:2014hha,Hooper:2020otu,Perez-Gonzalez:2020vnz,Bernal:2022pue,JyotiDas:2021shi,Calabrese:2023key,Calabrese:2023bxz,Schmitz:2023pfy,Ghoshal:2023fno,Barman:2024slw}, GW generation~\cite{Papanikolaou:2020qtd,Domenech:2020ssp,Bhaumik:2022pil,Bhaumik:2022zdd,Ghoshal:2023sfa}, or the stability of the Standard Model (SM) Higgs potential~\cite{Burda:2016mou,Burda:2015isa,Hamaide:2023ayu}. A substantial population of PBHs could further alter cosmological history by causing an early matter-dominated era, which would have concluded once these black holes evaporated due to Hawking radiation, or by modifying cosmic expansion during extended eras while extended mass distributions of PBHs evaporate over time~\cite{Cheek:2022mmy, Dienes:2022zgd,Barrow:1991dn}.
Finally, let us note that the PBH themselves could be the DM that we observe in the Universe~\cite{Carr:2016drx,Clesse:2016vqa,Carr:2017jsz,Green:2020jor,Croon:2020ouk}.

Several studies have investigated the PBH role in leptogenesis and their impact on baryon asymmetry. Ref.~\cite{Perez-Gonzalez:2020vnz} showed that PBH-produced RHNs can decay to generate lepton asymmetry. However, the entropy injection from PBHs—mainly producing photons—can dilute this effect, especially for PBHs with masses over $\sim \mathcal{O}(10^3)$ g. This dilution can place intermediate-scale leptogenesis scenarios with heavy PBH under tension. Even heavier PBHs (with initial masses $\MPBHini \gtrsim 10^{5.5}$g) evaporate after the Electroweak (EW) phase transition (EWPT), injecting entropy without producing asymmetry, place stringent constraints on both PBHs and leptogenesis \cite{Calabrese:2023bxz,Calabrese:2023key}.  Ref.~\cite{Bernal:2022pue} demonstrated that very light PBHs ($\mathcal{O}(1)$ g) could expand the parameter space for high-scale leptogenesis up to the GUT scale. These PBHs evaporate at high temperatures, producing RHNs that generate a $B-L$ asymmetry when washout processes are suppressed. Additionally, Kerr black holes can drive non-thermal leptogenesis over a broad range of PBH masses, particularly when heavy RHNs ($\mathcal{O}(10^{12}$ GeV)) interact with a scalar field \cite{Ghoshal:2023fno}. Even the lightest PBHs ($\MPBHini \lesssim 10^3$ g) can contribute to leptogenesis through mechanisms like wash-in leptogenesis, highlighting PBHs' significant role in early Universe baryogenesis \cite{Schmitz:2023pfy}. 

The production of a DM candidate in the early Universe can be particularly sensitive to the abundance of light evaporating PBHs ~\cite{Cheek:2021cfe, Gondolo:2020uqv,Masina:2020xhk,Baldes:2020nuv}. In the absence of any non-gravitational interaction between DM and the SM, the evaporation of PBHs simply constitutes an additional contribution to the DM relic abundance today~\cite{Gondolo:2020uqv,Baldes:2020nuv}. However, in the presence of interactions between dark-sector and SM particles (e.g. via the exchange of a dark mediator), the size of this contribution can vary widely across the parameter space~\cite{Cheek:2021cfe}. In particular, the variety of energy scales present in the problem --- namely the Hawking temperature of the evaporating PBH, the plasma temperature around the black holes, and the dark-sector particle masses --- can dramatically affect the capacity of DM particles produced via Hawking evaporation to propagate freely through the cosmos when they are produced. In Ref.~\cite{Cheek:2021cfe}, such interactions were considered and demonstrated to be relevant in the context of {\em Freeze-Out} and {\em Freeze-In} scenarios of DM production. 

All these previous treatments of DM and baryon asymmetry generation from PBHs have considered the heating of the plasma from PBH evaporation on a global scale, focusing on how the average plasma temperature increases due to entropy injection in the Universe. 
However, recent studies have shown that the heating of the primordial plasma around PBHs is highly non-uniform, instead producing localised hot spots ~\cite{Das:2021wei,He:2022wwy,He:2024wvt}. 
Hawking radiation deposits energy locally around the black hole, primarily through the Landau-Pomeranchuk-Migdal (LPM) effect~\cite{Landau:1953gr,Landau:1953um,Migdal:1956tc}, which is then diffused to the surrounding plasma. This produces a specific temperature profile around the PBH that can exceed the average plasma temperature by some orders of magnitude.

\begin{figure}[t!]
 \centering
 \includegraphics[width=\linewidth]{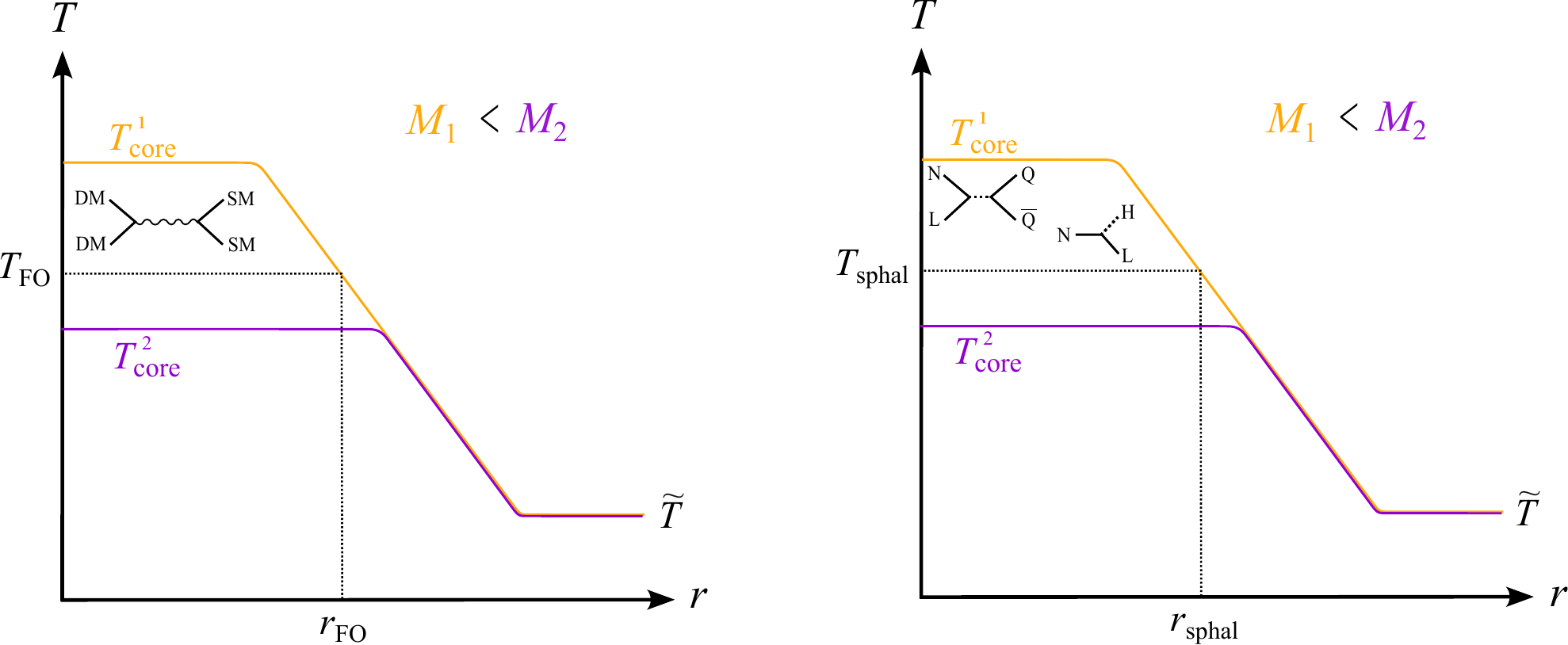}
 \caption{\label{fig:schem_hotspot} Schematic illustration of the effects of hot spots around PBHs on DM production and leptogenesis for two different PBH masses, $\MPBH_1$ (orange) and $\MPBH_2$ (purple), with $\MPBH_1<\MPBH_2$. Left: DM production. The radius $r_{\rm FO}$ represents the distance within the hot spot where the DM freeze-out temperature $T_{\rm FO}$ occurs. DM particles will thermalise for distances $r < r_{\rm FO}$. For the heavier PBH mass $M_2$, DM can escape the hot spot and contribute to the relic abundance since the plasma temperature is below $T_{\rm FO}$ everywhere. Right: Leptogenesis. The radius $\rsphal$ indicates the region where sphalerons remain active, with $T > \Tsphal$. Right-handed neutrinos decaying within $r < \rsphal$ will seed additional baryon asymmetry.}
\end{figure}
Here, we examine the impact of the presence of such hot spots on non-equilibrium processes, specifically leptogenesis and DM generation via freeze-out, as illustrated schematically in \figref{fig:schem_hotspot}. Interestingly, the presence of the hot spot influences these scenarios differently. 
For DM, if the hot spot's temperature exceeds the freeze-out temperature, DM particles may scatter off the heated plasma within the hot spot and thermalise, thereby erasing their contribution to the relic abundance. Only DM which escapes the hot spot contributes to the relic abundance. In contrast, if the hot spot's temperature exceeds the sphaleron freeze-out temperature for leptogenesis, only right-handed neutrinos that decay within the hot spot would contribute to the observed BAU. Those RHNs that escape the hot spots seed a leptonic asymmetry that is never converted to BAU.

This work is organised as follows. 
In Sec.~\ref{sec:PBHnut}, we describe PBH generalities in the early Universe, focusing on Hawking evaporation and the assumptions related to it, while simultaneously considering the main constraints to which this scenario is subject. 
Sec.~\ref{sec:PBH_HS} reviews the formation and evolution of hot spots around PBHs, following previous treatments. 
Then, we derive the probability for a PBH-produced gauge singlet particle to escape a hot spot in Sec.~\ref{sec:prob_escape}.
Finally, in Sec.~\ref{sec:applications} we apply the general formalism to the DM and leptogenesis scenarios to analyse the effect of the presence of hot spots in such cases.
We draw our conclusions in Sec.~\ref{sec:concl}, and we include App.~\ref{sec:AppA} where we detail the derivation of the thermally-averaged cross-section for particles with different temperatures and masses, and App.~\ref{app:AppB} where we give the relevant formulae for our DM model.
We consider natural units where $\hbar = c = k_{\rm B} = 1$, and define the Planck mass to be $M_{\rm PL}=1/\sqrt{G}$, with $G$ being the gravitational constant, throughout this manuscript.

\section{PBHs in the Early Universe} \label{sec:PBHnut}

The high density of the early Universe is a necessary but not sufficient condition for the formation of black holes with masses far below the Chandrasekhar limit, typically associated with astrophysical black hole formation.
Due to the high-pressure conditions in a radiation-dominated early Universe, the collapse of overdensities would be inhibited unless they surpass a specific threshold~\cite{Carr:2020gox,Carr:2020xqk}.
Many distinct scenarios have been proposed for the generation of such strong density perturbations, which will lead to the production of PBHs.
Given the plethora of mechanisms for black hole (BH) formation~\cite{Choudhury:2023jlt,Kawana:2021tde,Kawana:2022olo,Cotner:2019ykd,Baker:2021nyl,Baker:2021sno}, here we will assume that a PBH population emerged during an early radiation-dominated era from the collapse of primordial density perturbations.
We parameterise such a population through its initial mass, $\MPBHini$, and initial PBH energy density, $\beta^\prime$. The initial PBH mass, assuming a monochromatic mass distribution\footnote{Note that formation mechanisms, in general, predict extended mass distributions, see e.g.~\cite{Sasaki:2018dmp, Carr:2020xqk}.}, is related to the particle horizon via
\begin{align}\label{eq:initial_mass}
 \MPBHini &= \frac{4\pi}{3}\, \gamma\, \frac{\rhorad(\Tform)}{H^3(\Tform)}\sim 2\times10^7~{\rm g} \left(\frac{\gamma}{0.2}\right) \left(\frac{10^{12}~{\rm GeV}}{\Tform}\right)^2\,,
\end{align}
where $\gamma \sim 0.2$ is the gravitational collapse factor for a radiation-dominated Universe, $\rhorad$ is the radiation energy density with temperature $\Tform$ at BH formation and $H$ represents the Hubble rate. 
The initial PBH energy density $\rho_{\rm BH}(\Tform)$ is defined through its ratio to the total energy density $\rhotot$ at formation, modified to include the uncertainty of the gravitational collapse factor~\cite{Carr:2020gox},
\begin{align}\label{eq:beta}
 \beta^\prime \equiv \gamma^{1/2}\left(\frac{g_{\star, \rm form}}{106.75}\right)^{1/4}\frac{\rho_{\rm BH}(\Tform)}{\rhotot(\Tform)}\, = \gamma^{1/2}\left(\frac{g_{\star, \rm form}}{106.75}\right)^{1/4} \beta,
\end{align}
where $g_{\star, \rm form}$ are the relativistic degrees of freedom at formation.
Assuming the PBHs to be non-rotating and chargeless,  their initial Schwarzschild radius is
\begin{align}
 r_{\rm S, in} &= 2 G\MPBHini\sim 2.7 \times 10^{-8}~{\rm fm} \left(\frac{\MPBHini}{10^7~{\rm g}}\right)\,.
\end{align}
Therefore, we observe that the PBH population is made of microscopic black holes. Quantum effects become crucial for the evolution of these objects. 
Considering a semi-classical approach, Hawking~\cite{Hawking:1975vcx} demonstrated that black holes behave as black bodies, emitting all available degrees of freedom in nature with a temperature
\begin{align}\label{eq:HawkTemp}
 \TBH = \frac{1}{8\pi\, G \MPBH}\,\sim 10^3~{\rm TeV} \left(\frac{10^7 g }{\MPBH}\right).
\end{align}
Let us note that this radiation, known as Hawking radiation, begins once the surrounding plasma temperature cools to below the Hawking surface temperature of the PBH.
The differential emission rate per energy $E$, per time $t$, of a particle of type \X with mass $M_X$ is given by
\begin{equation}
 \frac{d^{2}N_X}{dt\, dE} = \frac{g_X}{2\pi}\frac{\vartheta(\MPBH,E)}{e^{E/T_{{\rm BH}}}-\delta_X}\,,\label{eq:emission_rate}
\end{equation}
where $\delta_X = (-1)^{2s_X}$ for \X with spin $s_X$ and internal degrees of freedom $g_X$. The spin-dependent absorption probability $\vartheta(\MPBH,E)$ represents the likelihood that a particle will reach spatial infinity, considering the effective potential barrier coming from the curved spacetime around the black hole. The PBH mass loss rate can be estimated using energy conservation arguments, and it is given by~\cite{Hawking:1975vcx, MacGibbon:1990zk, MacGibbon:1991tj}
\begin{align} \label{eq:MEq}
 \frac{d\MPBH}{dt} = -\sum_X \int_{M_X}^{\infty} \frac{d^{2}N_X}{dt\, dE}\, E\, dE = - \varepsilon(\MPBH)\, \frac{\Mpl^4}{\MPBH^2}\,,
\end{align}
where $\varepsilon(\MPBH)$ is the evaporation function that accounts for the degrees of freedom that can be produced at a given time, i.e., for an instantaneous mass. We will use the parametrisation of such a function following Ref.~\cite{Cheek:2021odj}.

The derivation of the Hawking spectrum is expected to hold until the black hole's curvature approaches the Planck scale, around $\ell_{\rm Pl}^{-2} \sim 10^{66}~{\rm cm^{-2}}$ ---where $\ell_{\rm Pl}$ is the Planck length--- which occurs when the black hole's mass approaches the Planck mass~\cite{Hawking:1975vcx}. However, Page~\cite{Page:1993wv,Page:2013dx} has shown that approximately halfway through its lifetime, at what is known as the Page time, the von Neumann entropy of the Hawking radiation exceeds the black hole's available degrees of freedom, which are typically assumed to equal the Bekenstein-Hawking entropy~\cite{Bekenstein:1972tm,Bekenstein:1973ur}. This leads to the well-known information paradox~\cite{Hawking:1976ra,Almheiri:2020cfm,Buoninfante:2021ijy}\footnote{Note that this issue arises only if the black hole's available degrees of freedom are assumed to be exactly equal to the Bekenstein-Hawking entropy; see Ref.~\cite{Almheiri:2020cfm,Buoninfante:2021ijy}}. Consequently, resolving these open questions might require us to revise our understanding of black hole time evolution or even question the validity of the semi-classical approximation after the Page time. 
Given the uncertainty about the extent of these potential changes, we assume that the semi-classical approximation remains valid up to approximately the Planck scale, so that the PBH time evolution follows the mass loss rate outlined in Eq.~\eqref{eq:MEq}.

The cosmological evolution of a Universe containing evaporating PBHs, SM radiation and other possible non-standard degrees of freedom, dubbed as ``DS'' from \emph{Dark Sector}, is governed by the Friedmann equations~\cite{Barrow:1991dn, Gutierrez:2017ibk, Cheek:2021cfe}
\be
\begin{aligned}\label{eq:FBEqs}
 \frac{d\varrho_{\rm SM}}{dN_e}  &= -\left.\frac{a}{H}\frac{d \ln\MPBH}{d t}\right|_{\rm SM}\varrho_{\rm PBH},~ \\
  \frac{d\varrho_{\rm DS}}{dN_e}& = -\left.\frac{a^{3w_{\rm DS}}}{H}\frac{d \ln\MPBH}{d t}\right|_{\rm DS}\varrho_{\rm PBH},~\\
  \frac{d\varrho_{\rm PBH}}{dN_e} & = \frac{1}{H}\frac{d \ln\MPBH}{d t}\varrho_{\rm PBH}\,,
\end{aligned}
\ee
where we have taken as time variable the number of e-folds, $N_e = \ln(a)$, with $a$ the scale factor. $\varrho_i$ are comoving energy densities ($\varrho_i = a^{3(1+w_i)}\rho_i$), so that the Hubble parameter $H$ is
\begin{align}
 H^2=\frac{8\pi}{3 M_{\rm Pl}^2}(a^{-4}\varrho_{\rm SM} + a^{-3(1+w_{\rm DS})}\rho_{\rm DS} + a^{-3}\rho_{\rm PBH}) \,,
\end{align}
and $w_{\rm DS}$ is the equation-of-state parameter of the Dark Sector\footnote{If the Dark Sector contains mixtures of two or more types of fluids, each component will have a different equation with their corresponding $w_{\rm DS}$.}.
Finally, in \equaref{eq:FBEqs}, the sub-index in $d\ln\MPBH/dt$ indicates that only the contribution to either the visible or dark sectors needs to be taken into account.
Since the PBH population behaves as a matter component of the Universe, there could be a PBH-dominated era, depending on the initial PBH energy density~\cite{Barrow:1991dn}.
This occurs when~\cite{Hooper:2019gtx,Lunardini:2019zob}
\begin{align}\label{eq:beta_crit}
  \beta^\prime \gtrsim \beta^\prime_{c} \equiv 2.5\times 10^{-14} \left(\frac{g_\star(\Tform)}{106.75}\right)^{-1/4}\left(\frac{M_{\rm BH}}{10^8~{\rm g}}\right)^{-1}.
\end{align}
Depending on whether and when the PBH population had evaporated in cosmological history, one can derive stringent constraints on the initial PBH fraction $\beta^\prime$~\cite{Carr:2020gox,Carr:2020xqk}.
PBHs with initial masses $\MPBHini\lesssim 10^8~{\rm g}$ would have evaporated before the BBN and the Cosmic Microwave Background (CMB) formation eras~\cite{Carr:2020gox,Carr:2020xqk,Keith:2020jww,Acharya:2020jbv,Boccia:2024nly}, and thus are subject to less severe or model-dependent constraints. 
For instance, Poissonian fluctuations in the number density of the PBH population can generate small-scale density variations, which subsequently lead to the generation of gravitational waves~\cite{Papanikolaou:2020qtd}. To ensure that the energy of these gravitational waves does not exceed that allowed at BBN we apply the following constraint~\cite{Domenech:2020ssp}
\begin{equation} \label{eq:GW}
 \beta \lesssim 1.1 \times 10^{-6} \left(\frac{\gamma}{0.2}\right)^{-\frac12} \left(\frac{\MPBHini}{10^4~\text{g}}\right)^{-\frac{17}{24}}.
\end{equation}
In what follows, we will concentrate on PBHs that would have evaporated prior to BBN, and thus have masses $\lesssim 10^8$~g.

\section{PBH Hot Spots}\label{sec:PBH_HS}

\noindent In this section, we explore the formation of hot spots around PBHs and how these hot spots evolve. In particular, we review the recent works of Refs.~\cite{He:2022wwy,He:2024wvt}, which provides the evolution of the local temperature around a single Schwarzschild PBH. 
In the early Universe, evaporating black holes heat the surrounding plasma, creating a distinctive temperature profile. This heating mechanism occurs as Hawking radiation diffuses through the thermal plasma \cite{Das:2021wei}. The local deposition of Hawking radiation occurs due to soft gauge radiation emission that occurs via multiple scatterings with the background plasma which is subject to the LPM effect \cite{Landau:1953gr,Landau:1953um,Migdal:1956tc}.
 Based on these results of \cite{He:2022wwy}, we will examine the evolution of the local temperature, \( T(r, N_e) \), around a single Schwarzschild PBH, where \( r \) denotes the radial distance from the Schwarzschild radius of the PBH.
 
\subsection{LPM Effect and Deposition of Energy} \label{sec:LPMsection}

\noindent Depending on the initial mass of the PBH, its associated Hawking temperature may be lower than that of the surrounding plasma, $\TBH < \Tplasma$ where $\Tplasma$ denotes the temperature of the plasma far from the PBH. In this case, no Hawking radiation is produced at all.
As the Universe cools, eventually $\TBH \gtrsim \Tplasma$ and the PBH begins to evaporate and the
Hawking radiation travels radially outwards. The typical momenta of these emitted particles is \( \langle \vec{p} \rangle \sim \TBH > \Tplasma \).
Initially, particles emitted by Hawking radiation are on their mass shell. However, as they scatter successively with the surrounding plasma, they can be lifted off their mass shells and subsequently radiate multiple low-energy (soft) daughter particles. This process causes the hard mother particle, originally emitted by the PBH, to eventually thermalise with the surrounding plasma by splitting into multiple daughter particles.

This thermalisation process is dominated by nearly collinear splitting of the Hawking radiation into soft daughter particles \cite{Harigaya:2014waa}, a process suppressed by the LPM effect. The LPM effect describes the suppression of bremsstrahlung radiation and pair production in a medium when the particle's energy is very high, due to multiple scattering within the medium. The LPM rate can be estimated as \cite{He:2022wwy}
\be
 \Gamma_{\rm LPM}(k,T) \sim \alpha^2 \frac{T^{\frac{3}{2}}}{\sqrt{|\vec{k}|}}\,,
\ee
where \( T \) denotes the local temperature of the plasma, \( |\vec{k}| \) is the modulus of the three-momentum of the daughter particle, and \( \alpha = g^2/(4\pi) \) is a fiducial value for the Standard Model fine structure constant, taken to be 0.1. This rate increases with both the coupling to the medium and the density of the medium. The hard parent particle, with initial energy \( E \), loses its energy and thermalises with the surrounding plasma at a rate given by \cite{He:2022wwy}
\[
\frac{1}{E}\frac{d E}{d t} \sim \frac{1}{E} \int^{E/2} \Gamma_{\rm{LPM}}(k,T) \, d k \propto \alpha^2 T \sqrt{\frac{T}{E}} \,,
\]
where the energy loss rate is largest when the daughter particle has momentum \( k \sim p/2 \).
Following the convention of Ref.~\cite{He:2022wwy}, we define
\[
 r_{\rm core}(\TBH,T) \equiv \Gamma_{\rm LPM}^{-1}(\TBH,T)\,,
\]
as the radius at which Hawking radiation, or equivalently the hard parent particle, produced by a PBH with temperature \( \TBH \), will thermalise with the plasma of temperature \( T \).
\begin{figure}[t!]
 \centering
 \includegraphics[width = 0.65\textwidth]{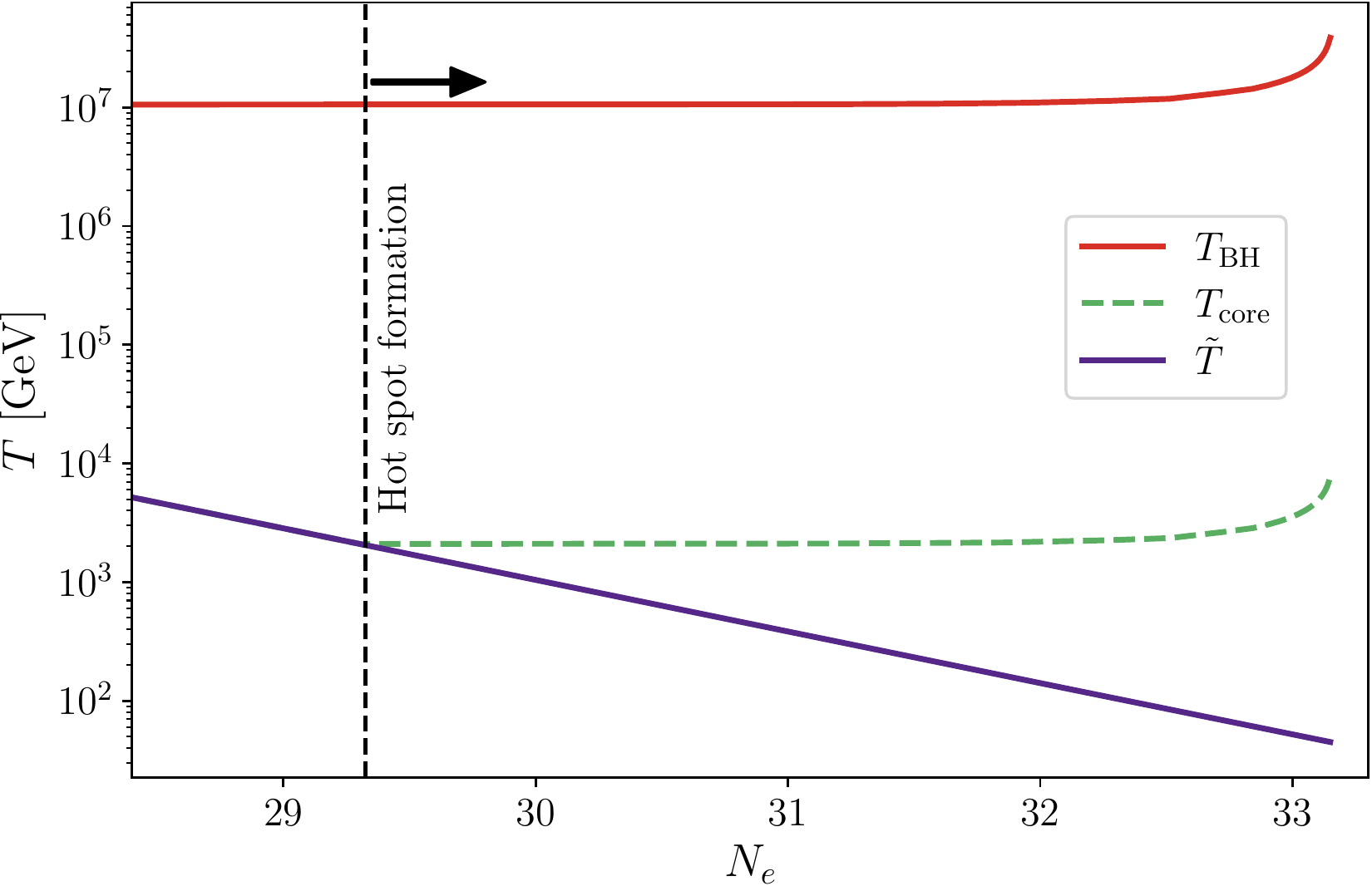}
 \caption{Comparison of the relevant temperatures involved in the formation and evolution of a PBH hot spot around a PBH with initial mass $\MPBHini = 10^6$g. The purple solid line indicates the background plasma temperature, the red solid line indicates the Hawking temperature of the BH and the green dashed line shows the temperature of the hot spot, diverging in temperature from the background at a well-defined moment.\label{fig:Tcomp}}
\end{figure}

\subsection{Hot Spot Profile}\label{sec:profile}

\noindent The energy deposited by Hawking radiation at \( r = r_{\rm core} \) diffuses both outwards and inwards. If the diffusion timescale over \( r_{\rm core} \) is shorter than the PBH evaporation timescale, the region \( r \leq r_{\rm core} \) becomes homogeneous. For a black hole with mass \( \MPBH \gtrsim 0.8\, \mathrm{g} \), diffusion is always efficient compared to evaporation, resulting in a constant core temperature, \( \Tcore \) \cite{He:2022wwy}. 
Taking the PBH as a heat source (see mass loss rate of \equaref{eq:MEq}) and using the conservation of energy density, one can show that the temperature of the homogeneous core at a given point in time is
\begin{equation}\label{eq:Tcore}
 \Tcore(N_e) = 2 \times 10^{-4} \left( \frac{\alpha}{0.1} \right)^{8/3} \left( \frac{g_{H*}} { g_*} \right)^{2/3} T_{\rm BH}(\MPBH(N_e))\,,
\end{equation}
where $g_*$ is the total relativistic degrees of freedom and $g_{H*}$ is the number of degrees of freedom of the Hawking radiation. We note that \equaref{eq:Tcore} describes instantaneous heating of the hot spot core by the energy deposited by the PBH.

From \equaref{eq:Tcore}, we observe that the core temperature of the would-be hot spot is approximately four orders of magnitude lower than the Hawking temperature. Therefore after the PBH starts to evaporate the entire Universe expands as normal (homogeneously) for another four decades in the scale factor. A temperature gradient arises between the locally heated region $r\leq r_{\rm core}$ and the rest of the Universe once $\Tplasma < \Tcore$. A comparison between the plasma, hot spot and Hawking temperature is shown in \figref{fig:Tcomp} for case where the initial black hole mass is $\MPBHini = 10^{6}$ g. The solid purple line shows the SM plasma's temperature evolution far from the PBH, which decreases linearly with the log of the scale factor. The Hawking temperature, which can be obtained from \equaref{eq:HawkTemp}, is shown by the solid red line and the dashed green line shows the hot spot temperature. Once the hot spot develops, the region $r \leq r_{\rm core}$, remains at an approximately constant temperature we denote as $\Tcore^{\rm ini}$. In the example shown in \figref{fig:Tcomp}, the hot spot develops at 
$N_e \sim 29$, with a temperature $\Tcore^{\rm ini} \approx T_{H}/10^{4}$. We note that towards the end of the PBH's lifetime, it can rapidly lose mass and this causes the Hawking temperature to rise. Consequently the hot spot temperature also rises. Once the hot spot has formed, but before PBH evaporation occurs, a non-trivial temperature profile around the hot spot develops. The critical radius, $\rcrit$\footnote{We note that we use the notation $\rcrit$ which is equivalent to the quantity $r_{\rm cr}$ used in \cite{He:2022wwy}}, is defined as the radius from the black hole over which the temperature profile, $T(r)$, is constant at $T_{\rm core}$~\cite{He:2022wwy},
\be
\label{rcrit}
 \rcrit \approx 6\times 10^7 \left(\frac{\alpha}{0.1}\right)^{-6}\left(\frac{g_*}{g_{H*}}\right)T_{\rm BH}^{-1}\,.
\ee
Another relevant length scale to model the evolution of the hot spot is the maximal length that thermal diffusion can smooth the hot spot within the evaporation time scale,
\be
 \rdec \approx 4.56\times 10^{14} \left[\frac{\alpha}{0.1}\right]^{-8/5}\left[\frac{g_*}{106.75}\right]^{1/5}\left[\frac{g_{H*}}{106.75}\right]^{-4/5} \left[\frac{T_{\rm BH}}{10^4\rm GeV}\right]^{-11/5}\,,
\ee
where $\rdec^{\rm ini}$ is defined to be $\rdec(\MPBH^{\rm ini})$ and is the longest length scale over which diffusion can smooth during the whole lifetime of the PBH. Radii $r > \rdec$ may have once been in thermal contact with the PBH, but freeze out at late times when the evaporation rate exceeds the diffusion rate over these larger length scales. As the PBH evaporates, $\rdec$ decreases and for radii $\rdec < r < \rdec^{\rm ini}$ the plasma once in thermal contact with the PBH freezes out into an envelope with temperature given by \cite{He:2022wwy}
\begin{equation}
     T_{\rm envelope} = 0.4\,\mathrm{MeV}\left(\frac{\alpha}{0.1}\right)^{6/5}\left(\frac{g_*}{106.75}\right)^{-2/5}\left(\frac{g_{H*}}{106.75}\right)^{3/5}\left(\frac{T_{\rm BH}^{\rm ini}}{10^4 \rm GeV}\right)^{7/5}\,.
\end{equation}
The temperature of the envelope falls with $r^{-7/11}$ until the background temperature of the Universe $\Tplasma$ is reached. Given these considerations, the temperature profile around a PBH following hot spot formation can be expressed as \cite{He:2022wwy}
\begin{equation}\label{eq:tprof}
T(r) = 
\begin{cases}
 \Tcore & r_{\rm BH} < r < \rcrit \\
 \mathrm{max}\left[\Tplasma, \Tcore \left(\frac{r}{\rcrit} \right)^{-1/3}\right] & \rcrit < r < \rdec\\
 \mathrm{max}\left[\Tplasma,T_{\rm envelope} \left(\frac{\rdec^{\rm ini}}{r}\right)^{7/11}\right] & \rdec < r < \rdec^{\rm ini} \\
 \Tplasma & r > \rdec^{\rm ini}
\end{cases}
\end{equation} 
where $\rcore, \Tcore$ and $\rdec$ are all implicitly time ($N_e$) dependent through their dependence on $\TBH$ which increases as the PBH evaporates, and $\Tplasma$ is explicitly a function of $N_e$ being the solution to the Friedmann equations in the background Universe. We note that we take $r_{\rm BH} = 5r_{S}$; however, the precise value of $r_{\rm BH}$ does not affect our results significantly as the hot spot is many times larger than the Schwarzschild radius.

\subsection{Hot Spot Separation and Evolution} \label{sec:PBH_sep}
In the absence of other nearby PBHs, the expression of the hot spot temperature exhibited in Eq.~\eqref{eq:tprof} is valid for any value of the radius. However, assuming that PBHs formed from the collapse of a Gaussian distribution of overdensities, PBHs are expected to spread in space in a Poissonian manner~\cite{Chisholm:2005vm} and are separated on average by a distance
\be
d_{\rm PBH} = \left(\frac{90 \MPBH}{4\pi^3 \beta g_\star(T_{\rm ev})\Tevap^3\Tform}\right)^{1/3}\,,
\ee
at the time when they evaporate. For a given temperature $\Tplasma < T < \Tcore$ (corresponding to a radius $r(T)$ after inverting Eq.~\eqref{eq:tprof}), one may wonder whether regions with temperatures larger than $T$ of different hot spots are expected to overlap when PBHs evaporate. In the next sections, it will be particularly important to know whether regions where sphaleron processes are still active or regions where DM particles are in thermal equilibrium with the plasma can be considered isolated regions of spacetime or whether hot spots are so close to each other that they merge. To answer this question, one may evaluate the ratio $2 r(T)/d_{\rm PBH}$. If this ratio is $\ll 1$, then hot spots can be safely assumed to be far away from each other when PBHs evaporate. In the opposite case, the merging of PBH hot spots should be taken into consideration, which is beyond the scope of this paper. In what follows, the region of parameter space corresponding to this case will be clearly identified.

Note that after evaporation is over, hot spots are not static but instead will evolve according to a diffusion equation that describes their evolution throughout cosmic history~\cite{He:2024wvt, heurtier}, growing with time until they merge to form a homogeneous Universe. For simplicity, in what follows, we will assume that after the PBH has evaporated any particle abundance produced in the vicinity of these hot spots becomes rapidly homogeneous in space. Given that the processes we are studying throughout this paper take place at a very early time in cosmology, such an assumption is reasonable, as hot spots have plenty of time to completely thermalise with their environment. 

\section{Can Hawking Radiation Escape Hot Spots?}\label{sec:prob_escape}
\noindent In this section, we consider the propagation of a generic Hawking radiation particle, $X$, travelling through the non-homogeneous local hot spot which has a temperature profile given by \equaref{eq:tprof}. We derive the mean free path for particle \X to scatter with the plasma within the hot spot via $1\to 2$ and $2 \to 2$ processes and calculate the likelihood of \X to escape the hot spot. Hawking radiation, produced at the Schwarzschild radius, travels radially outward with average momentum $\langle \vec{p} \rangle \sim \TBH \gg T(r)$ once the hot spot has formed, see \equaref{eq:Tcore}. Eventually interactions with the surrounding plasma cause the boosted Hawking radiation to thermalise. As discussed in Sec.~\ref{sec:LPMsection}, the dominant mechanism of energy deposition is the LPM-suppressed collinear splittings for any particles with gauge couplings. However, PBHs radiate into all available degrees of freedom. We may anticipate that particles with no or very small gauge couplings may be expected to free-stream out of the hot spot while relatively strongly interacting particles may be trapped within the hot spot. In the following, we consider the probability such a gauge singlet particle can decay or scatter within the hot spot. 

\subsection{Probability of Escaping the Hot Spot}\label{sec:escape}
\noindent In a homogeneous environment, the mean free path of a particle \X is given by
\be
\lambda = \frac{1}{\Gamma_X}\,, \ee
where $\Gamma_X$ is the total interaction rate of \X with the thermal bath, summing all allowed diagrams. One can see immediately that the above definition of $\lambda$ is insufficient in the case of Hawking radiation travelling through a hot spot since the interaction rate varies with the local temperature and density of the surrounding plasma. To allow for this radial variation, the mean-free path becomes
\be
\lambda\left(T(r)\right) = \frac{1}{\Gamma_X\left(T(r)\right)}\,.
\ee
Nonetheless, a physically meaningful definition of the mean free path should not depend on $r$. To address this, we define the probability for 
\X to reach a radius $r$ before interacting as,
\begin{equation}\label{P}
P(r) = e^{-\int^r_0\Gamma_X(r^\prime) \mathrm{d}r^\prime }\,,
\end{equation}
and set
\begin{equation}
 P\left( \lambda \right) \equiv \frac{1}{2}\,,
\end{equation}
such that the mean free path in a hot spot can be understood as the radius beyond which it is more likely than not that \X will have undergone some interaction. This definition of $\lambda$ allows us to quantitatively estimate the average distance travelled by Hawking radiation for a specific hot spot profile $T(r)$, which is calculated as in Sec.~\ref{sec:profile}. However, since the profile of any hot spot evolves with $N_e$, this measure does not provide a complete picture at all times. In particular, it will be of interest to calculate the comoving number density of \X produced by a PBH which escapes ($\mathcal{N}_X^{\rm escape}(r)$) to or is trapped within ($\mathcal{N}_X^{\rm trap}(r)$) radius $r$, as a function of time. These quantities are found by solving the following set of equations,
\be\label{eq:Nescape}
\frac{d \mathcal{N}_X^{\rm escape}(r)}{d N_e} = \frac{\Gamma_{\rm{PBH} \to X}}{H}\, P\left( r \right)\,\,\,,
\ee
\be\label{eq:Ntrapped}
\frac{d \mathcal{N}_X^{\rm trap}(r)}{d N_e} = \frac{\Gamma_{\rm{PBH} \to X}}{H}\,(1 - P\left( r \right) )\,\,\,,
\ee
respectively. $\Gamma_{\rm{PBH} \to X}$ is the instantaneous rate of production of particle \X by a PBH, found by integrating \equaref{eq:emission_rate} with respect to the particle energy. Integrating \equaref{eq:Nescape} (\equaref{eq:Ntrapped}) between the formation time of the hot spot, $\alphaform$, and evaporation time of the black hole, $N^{\rm evap}_e$, captures the total number density of \X that escapes to (are trapped within) $r$ over the hot spot lifetime. Near the end of the PBH lifetime, the Hawking temperature rapidly increases during the explosive final stages of the evaporation. Whether or not a black hole may radiate away more than $\sim 1/2$ of its initial mass is subject to debate in the literature \cite{Almheiri:2020cfm,Buoninfante:2021ijy,Juarez-Aubry:2023kvl,Burman:2023kko,Vachaspati:2006ki,Dvali:2020wft,Dvali:2024hsb}. In this work, we remain agnostic and take $\MPBH(N^{\rm evap}_e) \sim \mathcal{O}(100)\cdot\Mpl$.
The total interaction rate appearing in $P(r)$, $\Gamma_X(r)$, is a complicated function of $r$ because \X is moving through inhomogeneous plasma in which the temperature $T(r)$, particle comoving densities $\mathcal{N}(r)$, cross sections $\sigma(r)$, and degrees of freedom $g_*(r)$, all depend on the radial coordinate, $r$. $\Gamma_X(r)$ is given by
\be
\Gamma_X(r) = \Gamma_S(r) + \Gamma_D(r)\,, \ee
where $\Gamma_S$ sums all rates of 
arising from scattering of $\X$ with particles in the plasma while $\Gamma_D$ describes all available decay modes of $\X$.

\subsection{Decays in the Hot Spot}

\noindent We consider a generic particle, denoted as $X$, with mass $M_X$, which can undergo a two-body decay into particles $Y$ and $Z$. While Sec.~\ref{sec:leptogenesis} will focus on the specific case of right-handed neutrino (RHN) decay, our discussion here remains general. The particle \X is produced via Hawking radiation, assuming $\TBH \gtrsim M_{X}$, following an approximate blackbody spectrum. If $T_{\rm BH} \gg M_{X}$, the \X particles can be highly boosted. Consequently, the vacuum decay rate, $\Gamma_{X \to YZ}$, must be adjusted for time dilation, thermally averaged over the BH’s instantaneous spectrum, and it is given by
\begin{equation}\label{eq:tddecay}
 \Gamma_{X \to YZ} \left< \frac{M_{X}}{\TBH} \right>_{\rm BH} \approx \Gamma_{X \to YZ}\frac{K_1(z_{\rm BH})}{K_2(z_{\rm BH})}\,,
\end{equation}
where $z_{\rm BH} = M_{X}/T_{\rm BH}$ and we have assumed Maxwell-Boltzmann statistics. In our results, we do not use the analytical approximation from \equaref{eq:tddecay}; instead, we apply a full numerical solution that averages over the Hawking spectrum and accounts for the quantum statistics of the decay particles.
The vacuum decay rate $\Gamma_{X \to YZ}$ is kinematically unsuppressed if the mass of \X is greater than or equal to the sum of the masses of the final states. If the bare masses of any of these particles are small relative to the temperature of the thermal bath, thermal corrections on the order of $M^T \sim \alpha T$ will be significant. Since the temperature in the hot spot varies radially, the thermal masses of the particles will also vary accordingly. Therefore at radius $r$ the decay is kinematically accessible if the condition $ M_X^T(r) \geq M_Y^T(r) + M_Z^T(r)$ is satisfied, where $M^T$ indicates the thermally corrected mass. Where the kinematics vary radially, $\Gamma_X^{\rm BH}$ must be a function of radius and the full decay rate for \X in a hot spot should be given by
\be \label{GammaD}
\Gamma_D(r) = \left< \frac{M_{X}}{\TBH} \right>_{\rm BH} \Gamma_X^T(r) \,,
\ee
where $\Gamma_X^T$ is the rest frame decay width of \X including the radially-varying thermal corrections. 

\subsection{Scattering in the Hot Spot}
\noindent Hawking radiation may also scatter off the plasma as it travels through the hot spot. Considering only $2\to 2$ processes, the total rate of scattering of a PBH-emitted particle, $X$, with temperature $T_{\rm BH}$ and a particle drawn from the hot spot thermal bath with temperature $T(r)$ and mass $m_{j}$ is given by
\begin{equation}\label{GammaS}
\Gamma_S = \sum_\sigma n_j\langle \sigma \cdot v_{\text{Mol}} \rangle_{T_{\rm BH}T(r)} \,,
\end{equation}
where the summation is over all allowed scattering diagrams, $\langle \sigma \cdot v_{\text{Mol}} \rangle_{T_{\rm BH}T(r)}$ indicates the thermally averaged cross-section for the process $X,j \to k,l$, and $n_j(r)$ is the local number density of the particle species off which \X scatters. We emphasise that the local density of particle $j$ in the hot spot, $n_j(r)$, will vary as a function of radial distance since the local temperature varies. For instance, the hot spot temperature is largest within the ``core'' and in this region, $r<r_{\rm core}$, the number density of SM particles will be larger than in the cooler regions, $r> r_{\rm core}$. 
Further, we assume that particle $j$ is in thermal equilibrium within the hot spot.
In general, the scenario $M_{X}\neq m_{j}$ and $T_{\rm BH}\neq T(r)$ can arise and the thermally averaged cross-section is
\begin{equation}\label{eq:thermalAverage}
    \langle \sigma \cdot v_{\text{Mol}} \rangle_{T_{\rm BH}T(r)} = B \int^{\infty}_{s_\text{low}} \sigma(s) F \left ( \frac{C_1}{D}e^{-x_+^{\rm min}}(1 + x_+^{\rm min}) + \frac{\sqrt{C_2}}{\sqrt{D}}K_1 \left( \frac{\sqrt{D}}{T_{\rm BH}T(r)} \right) \right) \, ds\,,
\end{equation}
where $s$ denotes the centre-of-mass energy between particle \X and $j$, the lower integration boundary is $s_\text{low}=\max\left[\left(M_X + m_j\right)^2,\left(m_k + m_l\right)^2\right]$ $m_{k,l}$ are the masses of the outgoing particles and expressions for $B$, $x_{\rm min}^+$, $C_1$, $C_2$ and $D$ can be found in \appref{sec:AppA} \footnote{Note that compared to the general notation in \appref{sec:AppA} we make the identifications $T_1 = \TBH$, $T_2 = T(r)$, $m_1 = M_X$, $m_2 = m_j$. }. This result reduces to known results when $M_X = m_j$ \cite{Cheek:2021cfe} or $T_{\rm BH} = T(r)$ \cite{Gondolo:1990dk}. The masses appearing in \equaref{eq:thermalAverage} include any radially dependent thermal corrections. Since the specific hot spot profile, as well as the PBH temperature are functions of the scale factor, the thermal averaging of the cross sections evolves with $N_e$ in \equaref{eq:Nescape}. 
\section{Application to Leptogenesis and Thermal Dark Matter}\label{sec:applications}
\noindent In this section, we apply our considerations of the hot spot effect on Hawking radiation to two particle physics models and investigate the phenomenological implications. In Sec.~\ref{sec:leptogenesis}, we focus on the generation of the BAU through leptogenesis and how the PBH hot spots can affect this process. While in Sec.~\ref{sec:Zprime}, we consider the thermal production of a fermionic DM via a $Z^\prime$ portal.

\subsection{Leptogenesis}\label{sec:leptogenesis}
The origin of the BAU remains a mystery, but it is clear that the observed yield of the baryon asymmetry, 
$\YB^{\rm obs} \approx 8.7\times 10^{-10}$ \cite{Planck:2018yye},
should have existed before the onset of Big Bang Nucleosynthesis at $T_{\rm BBN} \approx 1$ MeV. Leptogenesis~\cite{Fukugita:1986hr}, which is based on the type-I seesaw mechanism \cite{Minkowski:1977sc, Yanagida:1979as, GellMann:1980vs, Mohapatra:1979ia}, explains the observed BAU by generating a net leptonic asymmetry, $\Delta L$, through the out-of-equilibrium dynamics of at least two gauge singlet fermions, also known as Right Handed Neutrinos (RHNs), $N_i$ where $i$ is a generational index. This leptonic asymmetry is subsequently converted into a baryonic asymmetry, $\Delta B$, by the $B+L$ violating EW sphaleron processes which are efficient at converting $\Delta L$ into $\Delta B$ until just after the EWPT when they freeze out at $\Tsphal \approx \Tewpt$. In Ref.~\cite{Calabrese:2023bxz} it was shown that a period of PBH domination can alter $\Tsphal$ in the background, in our analysis we are interested in the local freeze-out of sphalerons in each hot spot, so we always take $\Tsphal = 130$ GeV \cite{DOnofrio:2014rug}.
For $\MPBHini \gtrsim 10^{5.5}$g, the temperature of the background Universe is below $\Tsphal$ by the end of the PBH lifetime. 
Therefore, in principle, PBHs heavier than this value cannot produce any baryon asymmetry in the background through the production of RHNs because the sphaleron processes are exponentially suppressed. The most important effect of such heavy PBHs is to washout existing asymmetry through the production of a large number of photons \cite{Perez-Gonzalez:2020vnz,Calabrese:2023bxz,Calabrese:2023key}. 

Crucially, all of the previous work on the interplay between PBHs and leptogenesis assumes homogeneous heating of the Universe. However, inside a hot spot, sphalerons remain active where the local temperature $T(r)$ is heated above $\Tsphal$ by the thermalisation of the evaporation products of the PBH\footnote{Note that BHs act as seeds of baryon number violation, altering the sphaleron rate, when the Schwarzschild radius is of the same order as the electroweak scale, i.e., $\MPBHini\sim 10^{11}~{\rm g}$~\cite{DeLuca:2021oer}. However, for the BH masses that we are interested in, this effect is negligible.}. 
In this region, a leptonic asymmetry would continue to be converted into baryon asymmetry. We assume that after the PBH lifetime, the hot spot thermalises with the rest of the Universe, and that the conversion of lepton asymmetry into baryon asymmetry by sphalerons is instantaneous. We have numerically verified that the rate of washout processes that would erase the lepton asymmetry is always much smaller than the rate of outward diffusion within the hot spot, and we, therefore, neglect washout occurring in the hot spot. Therefore when RHNs produced by a PBH travel radially through its hot spot, their decays produce a lepton asymmetry which is converted into baryon asymmetry only if the decay occurs at $r \leq \rsphal$. The length scale $\rsphal$ is defined such that
\begin{equation}
 T(\rsphal) = \Tsphal\,,
\end{equation}
so that the number density of RHNs which decay within $r \leq \rsphal$ is $\mathcal{N}_{N}^{\rm trap}(\rsphal)$. By integrating \equaref{eq:Nescape} over the PBH lifetime, with $\Gamma(r)$ the total interaction rate for the RHNs, we can calculate $\mathcal{N}_{N}^{\rm trap}(\rsphal)$ and predict the corresponding contribution to $\YB$.

 The total scattering rate for $N_i$, $\Gamma_S(r)$, is given by \equaref{GammaS} where the sum over $\sigma$ includes all possible $2\to 2$ diagrams with $N_i$ in the initial state. Using the expressions given in \cite{Pilaftsis:2003gt} which include thermal corrections, we find that $\Gamma_S \ll \Gamma_D$ inside the hot spot and therefore neglect scattering such that $\Gamma(r) \approx \Gamma_D(r)$. $\Gamma_D(r)$ is the decay rate of the RHNs as a function of the radial distance from the PBH which is given by \cite{BoltzmannDecay,Giudice:2003jh}
\begin{eqnarray}
 \Gamma_{N_i}^T = \frac{M_{N_i}(Y^\dagger Y)_{ii}}{8\pi} \lambda^{\frac{1}{2}}(1,a_\phi,a_L) (1-a_\phi + a_L)\,\Theta (1-a_\phi - a_L)\,,
\end{eqnarray}
where $Y$ is the Yukawa matrix coupling the RHNs to the SM leptonic ($L$) and Higgs doublet ($\phi$), $\lambda(a,b,c) = (a-b-c)^2 - 4bc$, $a_\chi = (M_\chi/M_{N_i})^2$, and the Heaviside function reflects the kinematic restrictions. Depending on the local plasma temperature, $T(r)$, the lepton and Higgs thermal masses may suppress the decay rate of $N_i\to L \phi$. For the decay, $N_i \to L \phi$, to be kinematically accessible at $T(r) \geq \Tsphal$ we require roughly $M_{N_i} \gtrsim 0.1 \Tsphal$, where we take $0.1$ as a fiducial value representing the SM gauge couplings.
To calculate $\Gamma_D(r)$ the Yukawa matrix $Y$ must be known and for this purpose, we adopt the Casas-Ibarra parameterisation \cite{Casas:2001sr},
\begin{equation}
 Y = v_{\rm EW}^{-1} \sqrt{\MN} R \,\sqrt{\hat{m}_\nu} \, U_{\rm PMNS}^\dagger\,,
\end{equation}
where $v_{\rm EW} = 246$ GeV is the Higgs vacuum expectation value, $\MN$ is the (diagonal) mass matrix of the RHNs $\MN = {\rm diag}(M_{N_1},M_{N_2},M_{N_3})$, $R$ is a complex, orthogonal matrix which depends on three complex angles $\theta_{1},\theta_{2},\theta_{3}$, $\hat{m}_\nu$ is the (diagonal) mass matrix of the active neutrinos $\hat{m}_\nu = {\rm diag}(m_1,m_2,m_3)$, and $U_{\rm PMNS}$ is the Pontecorvo-Maki-Nakagawa-Sakata (PMNS) matrix \cite{Pontecorvo:1967fh,pontecorvo1957mesonium,pontecorvo1957inverse,Maki:1962mu}. The overall scale of the active neutrino mass matrix is not known but the mass splittings between the states are well measured. We assume normal ordering and fix the lightest neutrino mass to be $m_1 = 0$ eV such that $m_2 = \sqrt{\Delta m^2_{21}}$ and $m_3 \approx \sqrt{\Delta m^2_{31}}$. We use the best-fit values for the neutrino mass splittings and mixing parameters from \cite{Capozzi:2021fjo} (see also \cite{deSalas:2020pgw}). We consider a quasi-hierarchical texture of $\MN$ with the heavier two states $N_{2,3}$ nearly degenerate such that $M_{N_{3,2}} = \bar{M} \pm \Delta M/2$ where $\bar{M}$ is the common mass of $N_{2,3}$. The lighter state $N_1$ does not significantly contribute to $\YB$. This texture of $\MN$ is typical of the new Minimal Standard Model, $\nu $MSM, and has been studied for example in \cite{Calabrese:2023bxz,Klaric:2021cpi,Asaka:2005an,Granelli:2020ysj}. In this case, only one angle in the $R$ matrix is physically relevant, which we take as $\theta_{1} = x + iy$. The total mixing between the heavy and light states in our framework is given by \cite{Chianese:2018agp}
\begin{equation}\label{eq:Usq}
 U^2 = \frac{m_2-m_3}{2} \frac{\Delta M}{\bar{M}^2}\cos(2 x) + \frac{(m_2+m_3)}{\bar{M}} \cosh(2y) \,,
\end{equation}
which along with the common mass $\bar{M}$ forms the parameter space of experimental searches sensitive to RHNs \cite{Hirsch:2020klk,MATHUSLA:2022sze,Baldini:2021hfw,SHiP:2018xqw}. We choose points in parameter space which are testable at the ILC and CEPC experiments \cite{Yang:2023ice},
\be\label{eq:BM}
\begin{aligned}
 U^2 &= 10^{-5} \, , \, \bar{M} = 50\,\rm{GeV}\,, \\
 U^2 &= 10^{-7} \, , \, \bar{M} = 100\,\rm{GeV}\,,\\
 U^2 &= 10^{-4} \, , \, \bar{M} = 150\,\rm{GeV}\,.
\end{aligned}
\ee
\begin{figure}[t!] 
 \includegraphics[width = 0.49\textwidth]{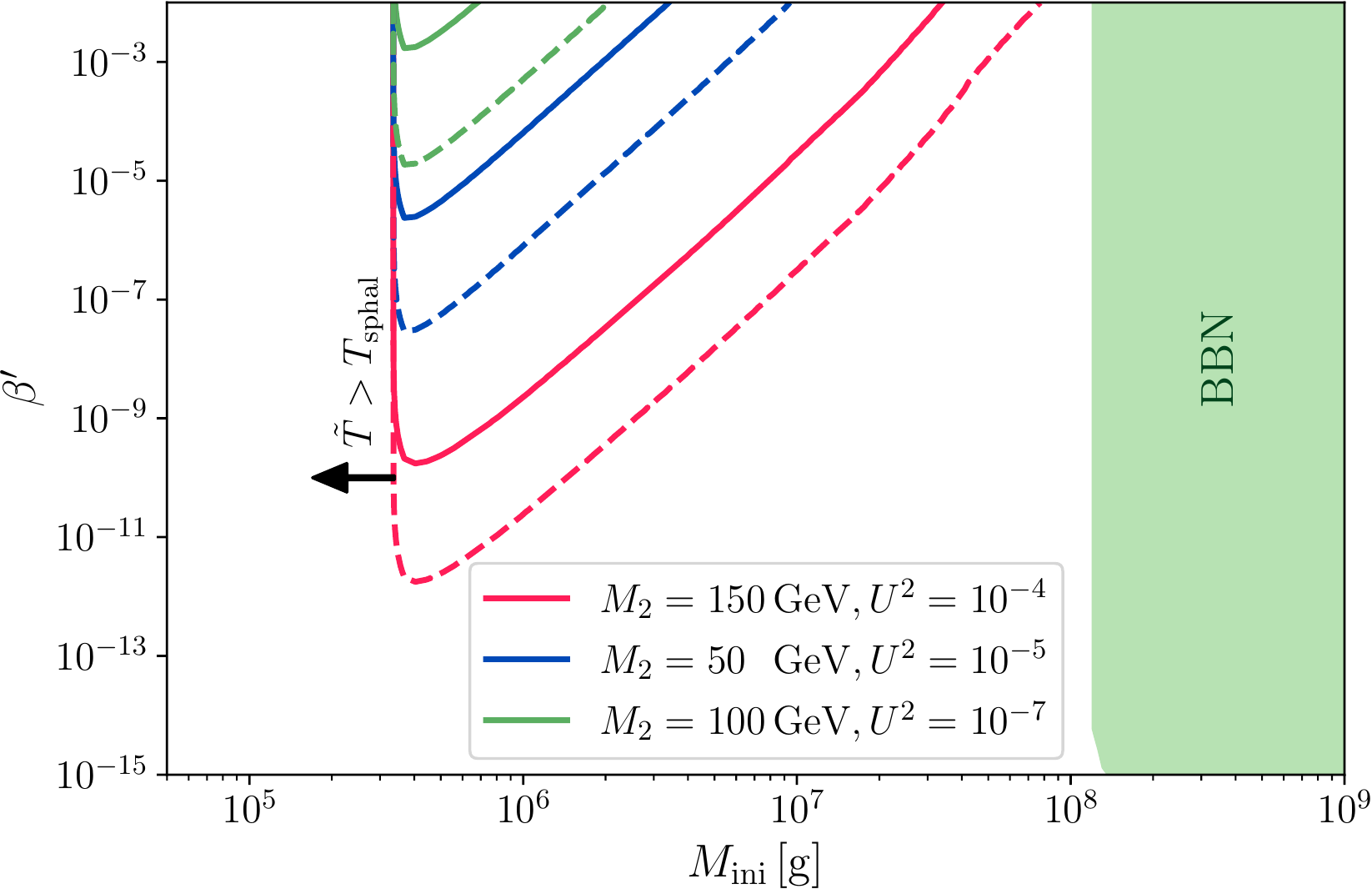}
 \includegraphics[width = 0.49\textwidth]{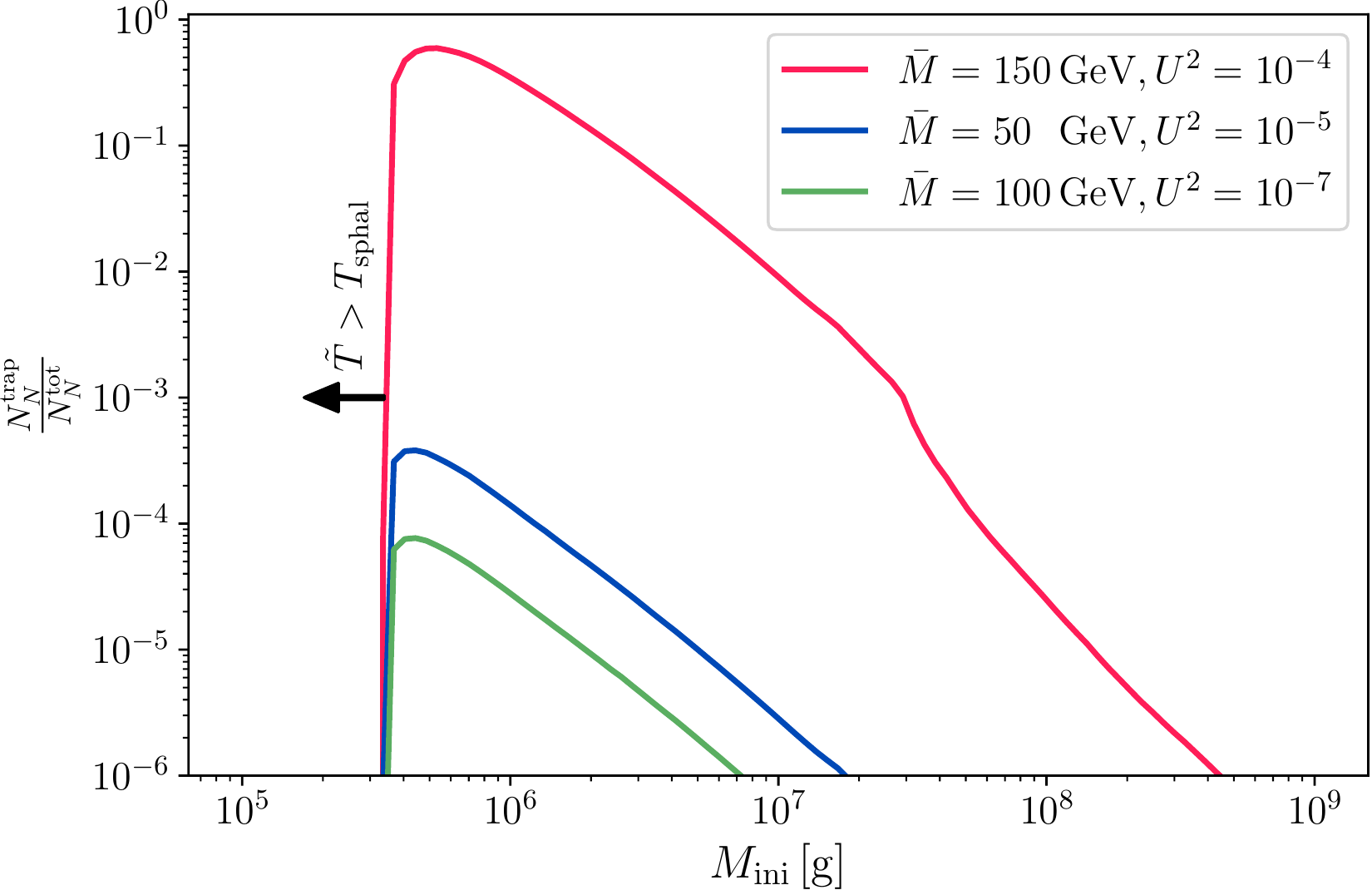}
 \caption{Left panel: Contours of $Y_B = Y_B^{\rm exp}$ for each benchmark point in \equaref{eq:BM}. The solid and dashed lines correspond respectively to $\Delta M/\bar{M} = 10^{-1}$ and $\Delta M/\bar{M} = 10^{-3}$. We report the most recent constraints on PBHs from BBN as a green shaded region \cite{Boccia:2024nly}. {The region to the left of the contours indicates where the plasma temperature is above the electroweak scale and therefore sphalerons are unsuppressed in the entire early Universe plasma.} Right panel: The fraction of RHNs produced by a PBH which are trapped within $\rsphal$ and therefore may contribute to $\YB$ is shown as a function of $\MPBHini$ for each benchmark point shown in \equaref{eq:BM}. In both panels, the lower limit in $\MPBHini$ occurs for $\Tevap = \Tsphal$, such that lighter PBHs never produce RHNs after the sphaleron freeze-out in the background Universe. \label{YBplot}}
\end{figure}
Due to strong washout effects, these benchmark points are incompatible with successful thermal leptogenesis in the two right-handed neutrino scenario. In our scenario, PBHs produce RHNs after the background plasma's temperature drops below \(\Tsphal\). As a result, any escaping RHNs that decay would not generate a baryon asymmetry. However, the area around PBHs can be heated above \(\Tsphal\), and depending on the coupling strength of the PBH-produced RHNs, they may decay within this hot spot where EW sphalerons are active. The resulting yield of baryon asymmetry can be estimated by solving the differential equation
\begin{equation}
\frac{d|\YB|}{d N_e} \approx \eta_{\rm sphal}\frac{|\epsilon|}{\mathcal{S}} \frac{d\mathcal{N}_{N}^{\rm trap}(\rsphal)}{d N_e}\NPBH\,,
\end{equation}
where $d\mathcal{N}_{N}^{\rm trap}(\rsphal)/d N_e$ is given by \equaref{eq:Ntrapped}, $\epsilon = \epsilon_2 + \epsilon_3$, with $\epsilon_i$ the CP violation parameter for $N_i$ summed over all flavours, $\NPBH$ is the comoving number density of PBHs, $\eta_{\rm sphal} \simeq 28/79 $ is the sphaleron conversion factor and $\mathcal{S}$ accounts for the entropy dump produced by the PBHs over their lifetime. As discussed earlier, the washout processes within the hot spot are inefficient compared to the outward diffusion and do not erase $\YB$.

In the left panel of \figref{YBplot} we show the contours of $\YB = \YBobs$ in the PBH parameter space ($\MPBHini,\beta^\prime)$, for the three benchmark points shown in \equaref{eq:BM} for two different relative mass splitting.  The green shaded region is the parameter space excluded by BBN \cite{Boccia:2024nly}. In the regions enclosed by the contours, $Y_{B} \geq\YBobs$ can be reproduced from decays of RHNs inside the sphaleron radius of hot spots, while no contribution to $\YB$ would be possible in the absence of hot spots. Remarkably, a baryon asymmetry is produced despite EW sphalerons having frozen out in the background Universe but remaining active in isolated hot spots. We observe that smaller $\Delta M$ (dashed lines in \figref{YBplot}) can produce the observed BAU for a smaller initial number density of PBHs ($i.e.$ smaller $\beta^\prime$) due to the relative enhancement in the decay asymmetry parameter, $\epsilon \propto 1/\Delta M$. It is also notable that only a mild ($\Delta M / \bar{M} \sim \mathcal{O}(10^{-1})$) mass degeneracy is required which can be detected given the anticipated level of experimental resolution \cite{SHiP:2021nfo}. This scenario may therefore be testable at the ILC and CEPC experiments \cite{Yang:2023ice}. 

The right-hand plot of \figref{YBplot} shows the fraction of RHNs produced by PBHs which are trapped within $\rsphal$ over the total number of PBH-produced RHNs as a function of $\MPBHini$. For smaller PBH masses, $\MPBHini \lesssim 10^{5.5}$ g, the whole universe remains hotter than $\Tsphal$, therefore $\mathcal{N}^{\rm trap}_N$ is undefined. Since the RHNs or their decay products cannot escape to a region where $\Tplasma < \Tsphal$, we expect the resulting baryon asymmetry to be washed out in this region. Lighter PBHs have hotter Hawking temperatures and therefore the PBH-produced RHNs are more boosted, compared to heavier PBHs, and their decay length is longer meaning they are more likely to escape the hot spot. Furthermore the size of the region $r < \rsphal$, is inversely proportional to the Hawking temperature (see \equaref{rcrit}) meaning more RHNs escape. As $\MPBHini$ increases, the initial temperature of the hot spot decreases so that for $\MPBHini \gtrsim 2\times10^7$g the hot spot forms at a temperature below $\Tsphal$. Only RHNs produced at late times when the core temperature of the hot spot increases can contribute to $\YB$ (see \figref{fig:Tcomp}) in this case. Thus for larger $\MPBHini$ the amount of time the hot spots are hotter than $\Tsphal$ decreases, trapping fewer RHNs within $r < \rsphal$. Further, the number of trapped RHNs increases with $U^2$ as increasing the mixing enhances the coupling of the RHNs with the plasma.

\subsection{Thermal Dark Matter}\label{sec:Zprime}
In this section, we examine how the hot spot around a PBH influences the DM production in the context of {\em Freeze-Out} and {\em Freeze-In} scenarios. We will estimate to what extent such effects alter the final relic abundance using a simple toy-model to support our conclusions.

\subsubsection{The Model}
\noindent In what follows, we will work in the framework of the so-called $Z^\prime$ portal DM models~\cite{Arcadi:2013qia,Chu:2013jja,Lebedev:2014bba,Bhattacharyya:2018evo,Mambrini:2010dq,Dudas:2012pb,Dudas:2013sia,Dudas:2009uq}. In particular, we will consider a massive Dirac fermion $\psi$ to be our DM candidate, singlet under the SM gauge group but charged under a dark Abelian symmetry $U(1)_D$. The gauge boson associated with this new symmetry is denoted as $Z'_\mu$. We assume that the mass of the latter originates from a St\"uckelberg mechanism \cite{Stueckelberg:1938hvi} so that we can safely ignore the existence of any other BSM particles in the spectrum of particles evaporated by the black hole. The Lagrangian we consider is
\bea\label{eq:lag}
\mathcal L &=& \mathcal L_{\rm SM}+\bar\psi(i\cancel{\partial}-m_{\rm DM})\psi
+\frac{1}{4}Z'_{\mu\nu}{Z'}^{\mu\nu}-\frac{1}{2}M_{Z'}^2 {Z'}_\mu {Z'}^\mu- g_{\rm D} {Z'}_\mu\bar\psi\gamma^\mu\psi- g_{\rm V} {Z'}_\mu\bar f\gamma^\mu f\,,\nonumber\\
\eea
where $f$ denotes any fermion in equilibrium with the SM bath, $m_{\rm DM}$ ($M_{Z^\prime}$) is the mass of the DM particle $\psi$ ($Z^\prime$), $Z_{\mu\nu}$ is the field strength tensor of $U(1)_D$ and $g_{\rm D}$ ($g_{\rm V}$) denotes the coupling of the $Z^\prime$ to the dark (visible) sector.

\subsubsection{Thermal Production}
\noindent The capacity of DM to thermalise with the SM plasma depends entirely on the values of the dark and visible couplings, $g_{\rm D}$ and $g_{\rm V}$. If these couplings are large enough, such that DM particles thermalise with the plasma before they become non-relativistic, the DM relic abundance will arise from DM particles {\em freezing out} of equilibrium at $T\lesssim m_{\rm DM}$ (the FO scenario). In the opposite case where these couplings are too small to establish a thermal equilibrium within the dark sector, then DM particles are slowly produced out-of-equilibrium from the thermal annihilation of SM particles and the DM relic abundance is said to {\em freeze in} (the FI scenario). In the absence of PBHs, the evolution of the relic density can be tracked numerically by solving the Boltzmann equation
\be
\dot{n}_{\rm DM} + 3H n_{\rm DM} = -\langle\sigma v \rangle_{\rm DM\to SM} ( n_{\rm DM}^2 - n_{\rm DM,\, eq}^2)\,,
\ee
where $n_{\rm DM}$ denotes the number density of DM particles, $n_{\rm DM,\, eq}=g_\star m_{\rm DM}^2 T/(2\pi^2)K_2(m_{\rm DM}/T)$ is the temperature-dependent number density of non-relativistic DM particles in thermal equilibrium, and $\langle \sigma v\rangle_{\rm DM\to SM}$ is the thermally averaged cross-section of DM annihilation.

In the presence of PBHs, this equation is modified by the inclusion of a source term corresponding to the injection of DM particles in the plasma from the Hawking evaporation of PBHs. In Ref.~\cite{Cheek:2021cfe}, the complete Boltzmann equations involving DM and SM particles, including PBH evaporation, were studied in detail. Whereas we do not intend to reproduce the results of Ref.~\cite{Cheek:2021cfe} in this paper, we will study to which extent the conclusions of this earlier work are affected by the presence of a hot spot around PBHs during their evaporation. In the two scenarios mentioned above, we use the following strategy:
\begin{enumerate}
 \item We adjust the parameters of the model such that DM is produced with the correct relic abundance in the absence of PBHs;
 \item Using these parameters, we calculate the annihilation and scattering cross section off the SM and DM particles in the plasma;
 \item Following the methodology detailed in Sec.~\ref{sec:escape} and given these cross sections, we calculate the total fraction of DM which is trapped by the hot spot;
 \item We estimate the region in the PBH parameter space $(\MPBHini,\beta^\prime)$ that would accordingly contribute to the relic density of DM at the $\mathcal O(1)$ level.

\end{enumerate}

\paragraph{\bf Freeze-In Production --}
In the case of FI production, the couplings $g_{\rm V}$ and $g_{\rm D}$ are typically very small, ensuring both that DM particles do not thermalise with the plasma and that the production of DM does not overclose the Universe. In this case, we find that the effect of the hot spot on the direct production of DM from PBH evaporation is always negligible. Indeed, the escape probability is always equal to one, and 100\% of the DM particles produced through Hawking radiation manage to escape the hot spot and propagate freely in the Universe without interacting with the plasma. The phenomenology described in Ref.~\cite{Cheek:2021cfe} regarding the FI case is thus robust and unaffected by the presence of PBH hot spots.

\paragraph{\bf Freeze-Out Production --}
We now consider the effect of the hot spot on the FO scenario, in which couplings can be $\mathcal O(1)$, causing ultra-relativistic DM particles produced by Hawking evaporation of PBHs to easily thermalise with the plasma outside the black hole. In the absence of PBHs, the FO of DM particles happens when the plasma temperature approaches the FO temperature, at which DM particles become non-relativistic and decouple from the thermal bath. In the presence of PBHs, this homogeneous decoupling is affected in two ways: $(i)$ PBHs produce DM particles via Hawking radiation and $(ii)$ these PBHs reheat the Universe, forming hot spots around them, which affects the time-evolution of the plasma temperature locally, hence the dynamics of the DM decoupling within these hot spots. The first aspect was already studied in previous works~\cite{Bernal:2020bjf, Bernal:2020ili, Bernal:2020kse, Bernal:2021bbv, Bernal:2021yyb, Bernal:2022oha, Bernal:2022pue, Masina:2020xhk, Masina:2021zpu, Gondolo:2020uqv, Baldes:2020nuv, Hooper:2019gtx, Cheek:2022dbx, Cheek:2022mmy}, in which the DM particles produced via Hawking radiation were assumed to either free-stream through the Universe, or to instantaneously thermalise with the plasma. However, throughout these references, the plasma surrounding PBHs was assumed to remain homogeneous even during reheating by PBHs. In the highly non-homogeneous environment of hot spots (see Sec. \ref{sec:PBH_HS}), DM produced as Hawking radiation may be trapped close to the PBH where the local temperature exceeds the freeze-out temperature $T_{\rm FO}$. This partly erases the contribution of PBH-produced DM, which one may naively think would free-stream and therefore contribute to the DM relic density if hot spots are not considered.

\begin{figure}[t!]
\centering
 \includegraphics[width = 0.65\textwidth]{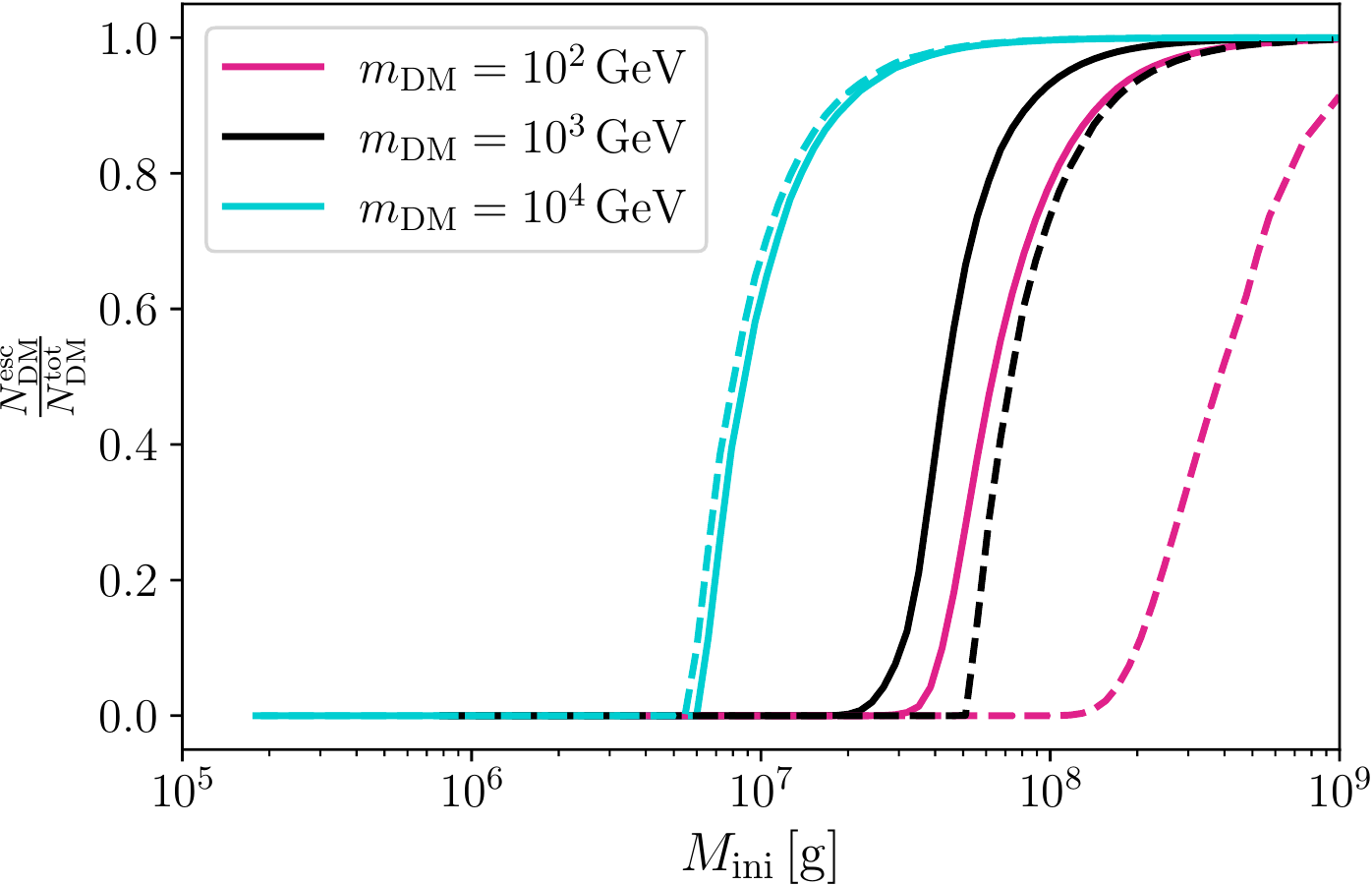}
 \caption{\label{fig:DMP} Characteristic curves showing the fraction of DM produced by a PBH which escapes to the freeze-out radius. The solid lines indicate  $M_{Z^\prime}/m_{\rm DM} = 10$ while the dashed lines indicate $M_{Z^\prime}/m_{\rm DM} = 10^{-3}$.}
\end{figure}

\begin{figure}[t!]
  \includegraphics[width = 0.49\textwidth]{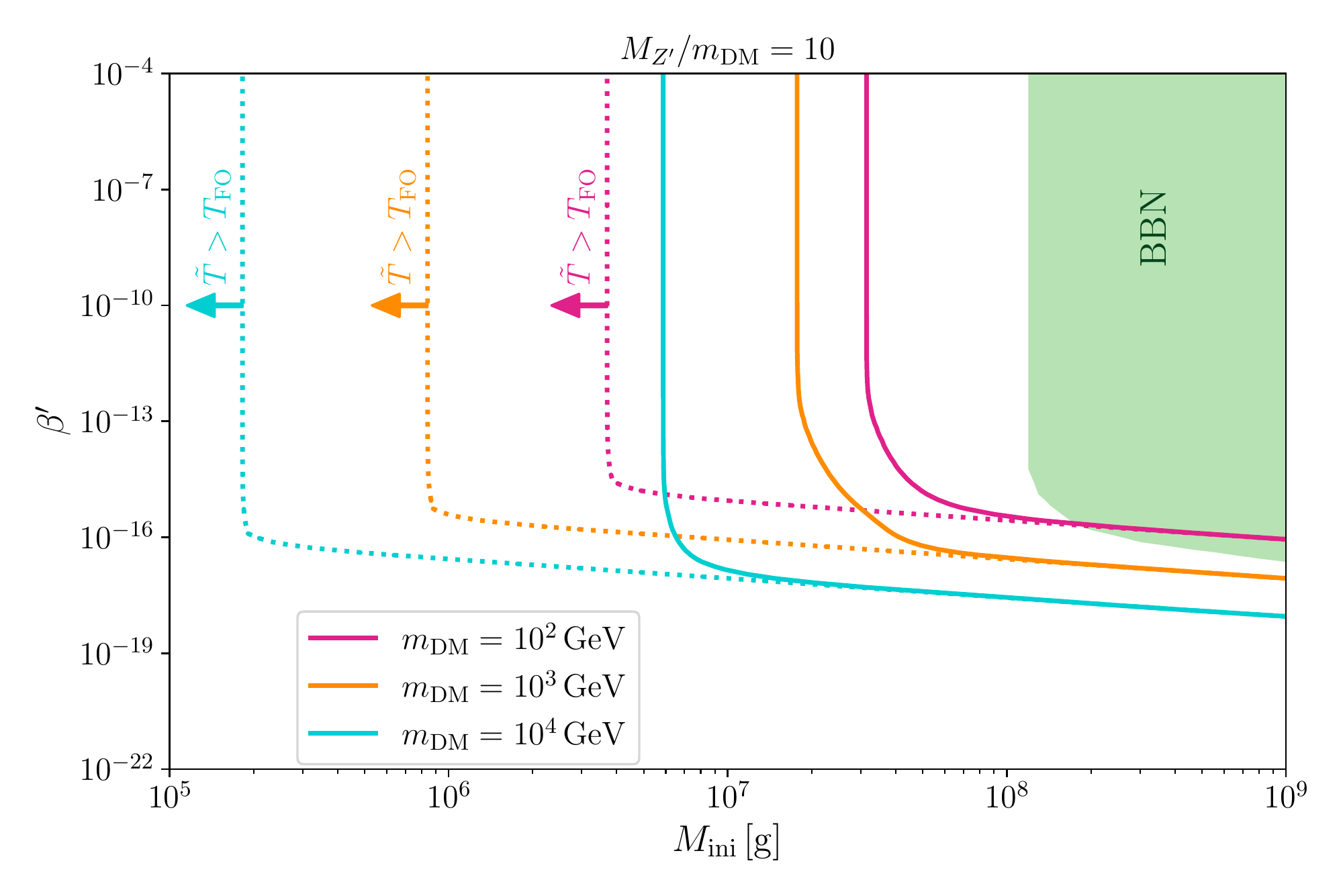}
 \includegraphics[width = 0.49\textwidth]{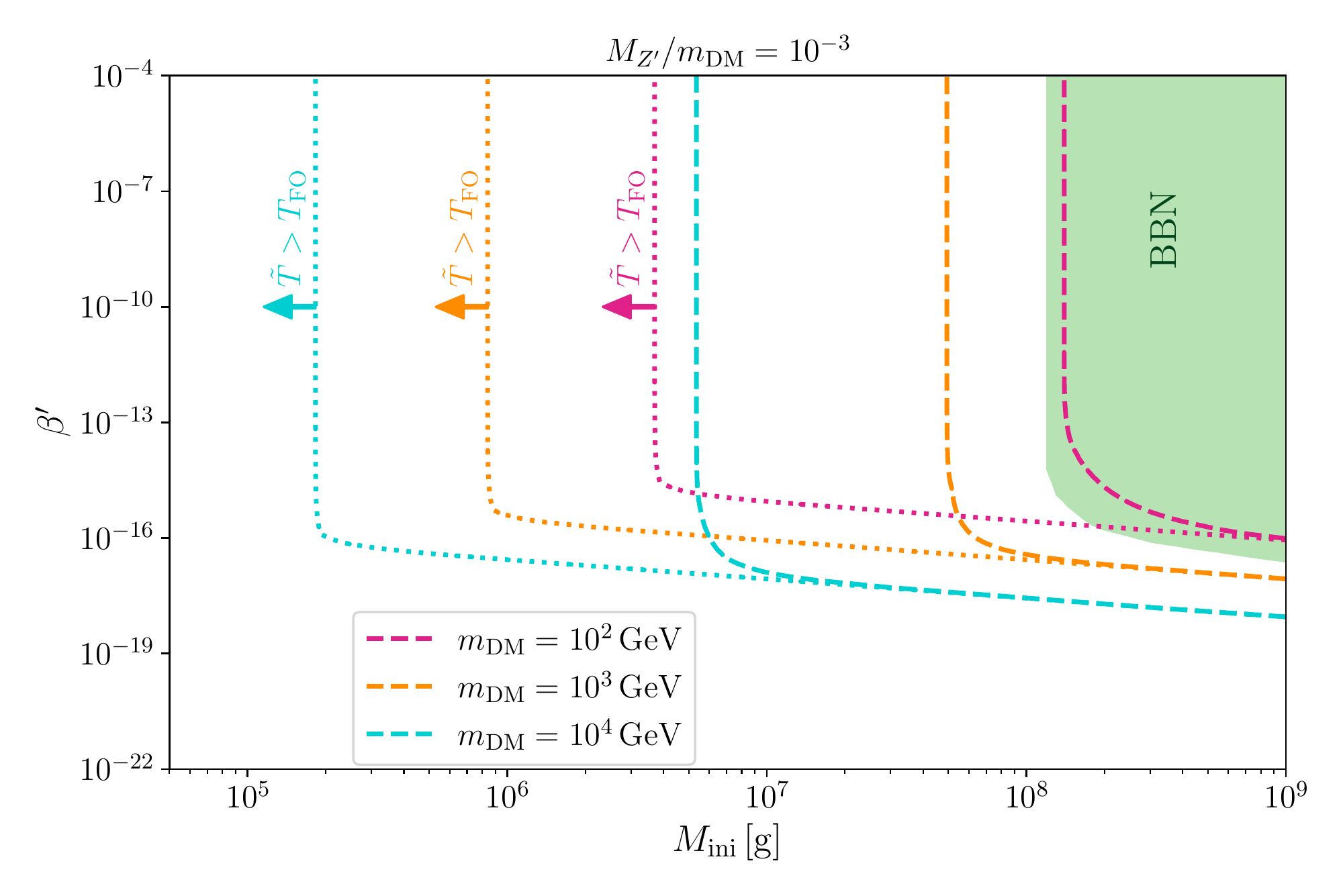}
 \caption{\label{fig:DM}Contours of $\Omega_{\rm DM}h^2= 0.12$ for three different masses of DM. In the left (right) panel, $M_{Z^\prime}/m_{\rm DM} = 10$ ($M_{Z^\prime}/m_{\rm DM} = 10^{-3}$). The dotted lines show the na\"ive calculation, that is without treating the hot spot properly whereas the solid (dashed) lines show the full calculation including the DM absorbing effect of the hot spots. In both panels, $g_{\rm D} = g_{\rm V} = g$ where $g$ is fixed such that the relic density of DM is reproduced via thermal freeze-out in the absence of PBHs. {In both plots, 
 the pink, orange and turquoise arrows show the region of the parameter space where the plasma temperature everywhere in the Universe exceeds the freezeout temperature. We show the most recent constraints on PBHs from BBN as a green shaded region \cite{Boccia:2024nly}.}}
\end{figure}

In \figref{fig:DMP}, we represent the ratio between the number of DM particles that manage to escape the hot spot with respect to the total number of DM particles emitted by each PBH, as a function of the PBH mass {$\MPBHini$} and for a few benchmark values of the DM mass. In that figure, the plain (dashed) lines correspond to the case of a heavy (light) $Z'$ mediator with $M_{Z'} = 10 m_{\rm DM}$ ($M_{Z'} = 10^{-3}m_{\rm DM}$). The DM that escapes the hot spot will contribute to the relic density, and therefore we find the effect of the hot spot is that for $\MPBHini \lesssim 10^{6.5}$ g the PBH-produced DM cannot contribute at all to the relic density.
Given this ratio, one can infer the PBH capacity to overclose the Universe, hence providing an upper limit on their abundance at formation, which is represented in \figref{fig:DM} for the same benchmark parameters, the left (right) panel corresponding to the case of a heavy (light) mediator. In both \figref{fig:DMP} and \figref{fig:DM}, we set the couplings $g_{\rm V}=g_{\rm D}$ such that DM is produced with the correct relic abundance in the plasma, far away from the hot spot. In \figref{fig:DM}, for comparison, we indicated with dotted lines the results when ignoring the presence of the hot spot. {We also indicated the region where the temperature of the Universe is larger than the FO temperature outside the hotspot. Similar to the case of leptogenesis, it turns out that the region of parameter space where $r_{\rm FO}>d_{\rm PBH}/2$, as explained in Sec.~\ref{sec:PBH_HS}, is contained within that region and PBH hot spots can safely be assumed to not overlap with each other during the evaporation.}

As one can see from \figref{fig:DM},  for heavy PBH masses, both cases (with and without hot spot) give similar results, as heavier PBHs feature colder hot spots that fail to trap DM particles and allow the PBH evaporation to contribute to the DM relic density freely. In the opposite case of lighter PBHs, the hot spot is hotter, and the scattering of DM particles is sufficient to trap these particles inside the hot spot and force them to thermalise instead of free streaming. The contribution of the PBHs to the DM relic abundance is thus reduced only to the thermal population of DM particles freezing out inside the hot spot when the latter cools down, which is comparable to the contribution from the DM freeze-out outside the hot spots. This effect is somewhat the opposite of the leptogenesis case, where RHNs were needed to decay within the sphaleron radius $\rsphal$ to be able to contribute to BAU, favouring light PBH masses. Here, the absorption of DM by the hot spot is significant for each DM mass considered, shifting the parameter space which overcloses the universe towards heavier $\MPBHini$.

The hot spots of lighter PBHs less efficiently absorb DM with larger masses. This is not intuitive, as heavy DM particles usually require larger couplings with SM particles to annihilate more efficiently during the FO process. However, it is important to note that, for a fixed PBH mass, heavier DM particles are produced with a smaller boost factor and that the region where the hot spot temperature exceeds the FO temperature --- of radius $r_{\rm FO} \equiv r(T_{\rm FO})$, as defined in Sec.~\ref{sec:PBH_HS} --- is also smaller in size. For these reasons, heavier DM particles escape more easily the hot spots of light PBHs, despite having stronger interactions with the SM plasma. For each choice of the DM mass, there exists a minimum PBH mass below which the temperature of the universe at the time of evaporation $\Tplasma$ exceeds the FO temperature. In that case, DM particles emitted by the black hole would always thermalise with the surrounding plasma, and the PBH cannot contribute to any excess in the DM relic density. This region is indicated in \figsref{fig:DM}{fig:ratios} by arrows pointing towards the corresponding region, explaining why the relic density parameter space closes, even in the absence of a hot spot.

A qualitative difference between the light and heavy mediator cases is that the DM annihilation (which is the main process controlling its freezing out from the plasma) is entirely given by the $s$-channel annihilation process $\bar \psi\psi \to Z'\to \bar f f$ in the heavy mediator case, but also by the $t$-channel annihilation process $\bar \psi \psi\to Z'Z'$ in the light mediator case. As can be seen from the formulas listed in Appendix~\ref{app:AppB}, the $s$-channel annihilation cross-section scales like $g_{\rm V}^2g_{\rm D}^2$ whereas the $t$-channel annihilation cross-section scales like $g_{\rm D}^4$. Therefore, in the light mediator case, the annihilation cross section can be mainly controlled by the dark coupling, while its interaction with the hot spot plasma can be suppressed in the case of a small visible coupling. 
\begin{figure}
    \centering
    \includegraphics[width=0.6\linewidth]{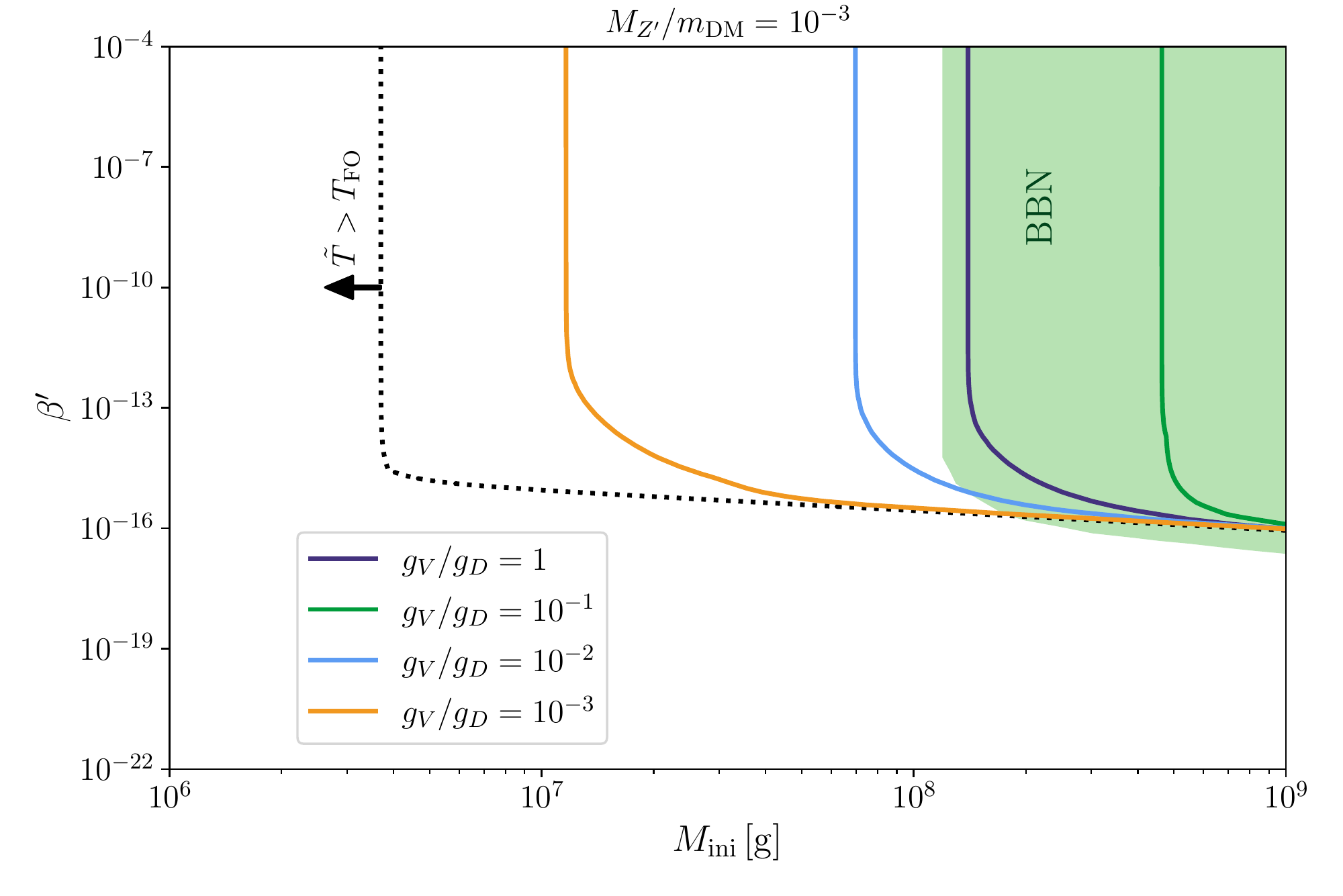}
    \caption{\label{fig:ratios}Same results as \figref{fig:DM}, but for $m_{\rm DM}= 10^3 M_{Z'}=100\, \mathrm{GeV}$ and various values of the ratio $g_{\rm V}/g_{\rm D} = 1$ (purple), $10^{-1}$(green), $10^{-2}$ (light-blue), and $10^{-3}$ (orange). The dotted line shows the calculation without considering the effect of hot spots.}
\end{figure}
In \figref{fig:ratios}, we vary the ratio between the visible and dark couplings in the light mediator case and show that the effect of the hot spot is reduced in this limit. Nonetheless, such a situation cannot be pushed to the limit where the visible coupling is exactly zero, as this would imply that the mediator is not maintained in equilibrium with the SM plasma and the dark sector would have its own equilibrium, leading to a different phenomenology~\cite{Heurtier:2019eou,Berlin:2016gtr,Berlin:2016vnh}. Therefore, it is expected that the effect of the hot spot is always significant in the case of a WIMP candidate since all the cases we surveyed show that hot spots tend to suppress the escape probability of DM for small PBH masses.

\section{Conclusion}\label{sec:concl}

\noindent In this work, we explored the role that primordial black hole (PBH) hot spots play in addressing two of the most pressing questions in cosmology: the origin of the baryon asymmetry of the Universe (BAU) and the nature of dark matter (DM). By properly accounting for the localised hot spots formed around PBHs due to Hawking radiation \cite{He:2022wwy,He:2024wvt}, we reveal deviations from previous studies that assumed homogeneous heating of the primordial plasma. For low-scale leptogenesis, our findings show that the presence of these hot spots can sustain electroweak sphaleron processes locally, even after they have frozen out in the broader Universe. This allows the lepton asymmetry produced by the decay of right-handed neutrinos within these hot spots to be converted into baryon asymmetry. Our mechanism enables the generation of the BAU in regions of leptogenesis parameter space where thermal leptogenesis would not be viable in standard cosmology.  Additionally, this mechanism allows PBHs, which could not produce baryon asymmetry under the homogeneous heating assumptions commonly used in the literature, to do so within the hot spot framework. Importantly, these regions of the parameter space tend to favour larger mixing of the RHNs with the active neutrinos and are in testable regions of the RHN parameter space. This underscores the importance of including the effects of hot spots in cosmological models.

In the context of thermal DM production, we analysed freeze-out and freeze-in scenarios in the presence of PBH hot spots. We find that in FI scenarios, the impact of hot spots is minimal, with DM particles produced via Hawking evaporation largely escaping and contributing to the relic density as previously estalished in the literature. 
For FO scenarios, since the temperature inside the hot spot can exceed the DM FO temperature, PBH-produced DM can be trapped in the hot inner region, and thus it can be efficiently thermalised with the plasma, erasing its contribution to the relic abundance.
Only the DM particles which travel far from the PBH and free stream in the Universe contribute to the observed abundance of DM today. This effect leads to large corrections in the regions of PBH parameter space which would overclose the Universe in DM scenarios.

Before we conclude, let us comment on a few aspects of this paper that would deserve further exploration in the future. First, we have assumed throughout this work that the production and scattering of BSM particles do not affect the time evolution of the hot spot temperature profile. In reality, the fact that particles may be emitted by the black hole and deposit their energy at distances that are larger than what is expected from SM particles may affect this evolution and the overall shape of the temperature profile. This would be particularly interesting in the case PBHs which would emit more than one dark sector particle~\cite{Baker:2022rkn, Calza:2021czr, Perez-Gonzalez:2023uoi, Calza:2023iqa, Calza:2023rjt}, in which case the morphology of the energy deposition throughout the hot spot could be significantly altered by the presence of BSM physics. Second, we have assumed that PBHs contribute to the BAU or the DM relic abundance in a homogeneous fashion following the equilibration of the plasma after the PBH lifetime. In reality, the thermal history of the Universe is locally affected by the presence of PBH hot spots. Whereas the distance that separates PBHs from one another when they evaporate may be much smaller than the Hubble radius at that time, --- which is commonly used to argue that PBHs inject particles in a homogeneous way in cosmology --- this modified evolution of the thermal bath on small scales could lead to spatial variation in the DM abundance, or the baryon/lepton asymmetry of the Universe, leading to observable signals. We leave the details of such a spatial distribution for future work. 

Our findings underscore the necessity of accounting for the non-uniform temperature profiles around PBHs when considering their role in early-Universe cosmology. In particular, we demonstrated that the interplay between PBHs, hot spots, and non-equilibrium processes like leptogenesis and DM production could provide new insights into the early Universe's conditions, potentially altering these scenarios' testability in future experiments. More generally, our findings suggest that evaporating PBHs may heat the Universe forming hot spots that may {\em screen} the emission of hot particles from evaporating PBHs on scales much smaller than the average distance between PBHs. This may significantly impact the way PBHs affect cosmology and particle physics beyond the sole cases of DM and baryon asymmetry production. 

\section*{Acknowledgements}
We want to thank the organisers and participants of the Focus Week on Primordial Black Holes 2023 where part of this work was presented.
The work of LH is supported by the STFC (grant No. ST/X000753/1). YFPG and JT are supported by the STFC under Grant No.~ST/T001011/1. JWG is grateful to IPPP, Durham University for their hospitality during the completion of this work. 

\begin{appendix}

\section{Cross Section Thermal Averaging for particles with different masses and temperatures}
\label{sec:AppA}

We consider in this appendix the thermal averaging of a $2\rightarrow 2$ scattering processes involving two particles, with masses $m_1, m_2$, having thermal distributions with different temperatures $T_1, T_2$. 
In this case, the velocity averaged cross section is defined as
\begin{equation} \label{int}
 \langle \sigma \cdot v_{Mol} \rangle_{T_{1} T_{2}} \equiv \frac{\int
 \sigma \cdot v_{Mol} f_1 f_2\, d^3p_1 \, d^3p_2}{\int d^3p_1 f_1 \, \cdot \int d^3p_2f_2}
\end{equation}
where $v_{Mol}$ is the Moller velocity, and we take both interacting particles to be Maxwell-Boltzmann distributed, i.e., $f_i = e^{-E_i/T_i}$. The volume element can be recast as 
\begin{equation}
 d^3p_1\,d^3p_2 = 2\pi E_1 E_2 \, dE_1 \,dE_2\, d \cos \theta
\end{equation}
where $\theta$ is the angle between the momentum vectors of the two incoming particles which have energy $E_{1,2}$. We change variables to 
\begin{eqnarray}
x_+ \equiv \frac{E_1}{T_1} + \frac{E_2}{T_2} \\
x_- \equiv \frac{E_1}{T_1} - \frac{E_2}{T_2} \\
s = (p_1^\mu + p_2^\mu)^2
\end{eqnarray}
where $p^\mu_i$ is the four-momentum of particle $i$. Calculating the Jacobian leads to 
\begin{equation}
d^3p_1\,d^3p_2 = 2\pi^2 T_1 T_2 E_1 E_2 dx_-\, dx_+ \, ds
\end{equation}
so that the numerator of the integral \equaref{int} becomes
\begin{equation}
 2\pi^2 T_1 T_2 \int \sigma \cdot v_{Mol} f_1 f_2 E_1 E_2 dx_- \, dx_+ \, ds = 2\pi^2 T_1 T_2 \int \sigma F e^{-x_+} dx_- \, dx_+ \, ds 
\end{equation}
where $F = v_{Mol}E_1 E_2$. The limits of integration transform as follows. We have
\begin{eqnarray}
 E_1 \geq m_1 \\
 E_2 \geq m_2 \\
 -1 \leq \cos \theta \leq 1
\end{eqnarray}
and since in general $m_1 \neq m_2$, we have
\begin{equation}
 s = m_1^2 + m_2^2 + 2E_1E_2 - 2|p_1||p_2|\cos \theta
\end{equation}
so that 
\begin{equation}
 s \geq m_1^2 + m_2^2 + 2E_1E_2 - 2|p_1||p_2|
\end{equation}
This inequality can be solved for $x_-^{\rm max}$, the result reads
\begin{equation}
 x_-^{\rm max} = \frac{ x_+ C_1 + \sqrt{C_2(T_1^2T_2^2x_+^2 - D})}{D}
\end{equation}
where we have made the identifications $C_1 = m_1^2T_2^2 - m_2^2T_1^2$, $C_2 = m_1^4 + \,(\,m_2^2 -s\,)\,^2 -2m_1^2\,(\,m_2^2 + s\,)$, and $D = sT_1T_2 + \,(\,T_2 - T_1\,)\,\,(\,m_1^2T_2 - m_2^2 T_1)$.
Clearly $x_-^{\rm min} = 0$. For $x_+$ the upper limit is of course $\infty$, while the lower limit can be obtained straightforwardly by requiring that $x_-^{\rm max}$ be real, so that
\begin{eqnarray}
 x_+ \geq \frac{\sqrt{D}}{T_1T_2}
\end{eqnarray}
and finally we have $s_{\rm min} = (m_1 + m_2)^2$.

Our numerator can now be calculated as
\begin{equation}
2\pi^2 T_1 T_2 \int^{\infty}_{s_{\rm min}} \sigma F \, ds \int^{\infty}_{x_+^{\rm min}} e^{x_+} \, dx_+ \int^{x_-^{\rm max}}_0 dx_- = 2\pi^2 T_1 T_2 \int^{\infty}_{s_{\rm min}} \sigma F \, ds \int^{\infty}_{x_+^{\rm min}} e^{x_+}x_-^{\rm max} \, dx_+ 
\end{equation}
First considering the integration over $x_+$, we can split the integral as
\begin{equation}
 \int e^{-x_+}x_-^{\rm max} \, dx_+ = \frac{C_1}{D}\int e^{-x_+}x_+ dx_+ + \frac{\sqrt{C_2}}{D}\int e^{-x_+}\sqrt{ T_1^2T_2^2x_+^2 - D}dx_+
\end{equation}
The first term gives
\begin{equation}
\frac{C_1}{D}\int e^{-x_+}x_+ dx_+ = \frac{C_1}{D}[-e^{-x_+}(1+x_+)]^{\infty}_{x_+^{\rm min}} = \frac{C_1}{D}e^{-x_+^{\rm min}}(1 + x_+^{\rm min})
\end{equation}
while for the second term we can define
\begin{align}
 u &\equiv \frac{T_1T_2}{\sqrt{D}}x_+\\
 du &= \frac{T_1T_2}{\sqrt{D}}dx_+\\
 u_{\rm min} &= 1
\end{align}
so that the integral becomes
\begin{equation}
 \frac{\sqrt{C_2}}{T_1T_2}\int^{\infty}_1 e^{-u \sqrt{D}/(T_1T_2)}\sqrt{u^2 - 1}du
\end{equation}
This integral can be evaluated by using the integral form of the modified bessel functions \cite{abramowitz+stegun}
\begin{equation}\label{Basset}
 K_n(z) = \frac{\sqrt{\pi}}{(n-1/2)!}\left(\frac{z}{2}\right)^n \int^\infty_1 e^{-zx}(x^2 - 1)^{n-1/2}dx
\end{equation}
so that by setting $n = 1$ and $z = \sqrt{D}/(T_1T_2)$ we obtain
\begin{equation}
 \frac{\sqrt{C_2}}{T_1T_2}\int^{\infty}_1 e^{-u \sqrt{D}/(T_1T_2)}\sqrt{u^2 - 1}du = \frac{2\Gamma\left(\frac{3}{2} \right)\sqrt{C_2}}{\sqrt{\pi D}}K_1 \left( \frac{\sqrt{D}}{T_1T_2} \right) = \frac{\sqrt{C_2}}{\sqrt{D}}K_1 \left( \frac{\sqrt{D}}{T_1T_2} \right)
\end{equation}
Therefore, the numerator of the thermally averaged cross section can be written as
\begin{equation}
 \int \sigma \cdot v_{mol} f_1f_2\, d^3p_1 \, d^3p_2 = 2\pi^2 T_1 T_2 \int^{\infty}_{(m_1 + m_2)^2} \sigma F \left ( \frac{C_1}{D}e^{-x_+^{\rm min}}(1 + x_+^{\rm min}) + \frac{\sqrt{C_2}}{\sqrt{D}}K_1 \left( \frac{\sqrt{D}}{T_1T_2} \right) \right) \, ds
\end{equation}
For $m_1 \neq m_2$, the correct expression for $F$ is $F = \frac{1}{2}\sqrt{(s-m_1^2-m_2^2)^2 - m_1^2m_2^2}$ \cite{Gondolo:1990dk}.
For the denominator, we have
\begin{equation}
 16\pi^2\int^\infty_{m_1} e^{-E_1/T_1} \sqrt{E_1^2-m_1^2} E_1 dE_1\int^\infty_{m_2} e^{-E_2/T_2}\sqrt{E_2^2-m_2^2} E_2 dE_2
\end{equation}
The $E_1$ integral gives
\begin{equation}
 m_1^2 T_1 K_2\left(\frac{m_1}{T_1}\right)
\end{equation}
so that the full result for the denominator is
\begin{equation}
 16\pi^2 T_1 T_2 m_1^2 m_2^2 K_2 \left( \frac{m_1}{T_1}\right) K_2 \left( \frac{m_2}{T_2} \right)
\end{equation}
Finally, we can write the thermal averaged cross section as
\begin{equation} \label{TA}
 \langle \sigma \cdot v_{Mol} \rangle_{T_1T_2} = X \int^{\infty}_{(m_1 + m_2)^2} \sigma(s) F \left ( \frac{C_1}{D}e^{-x_+^{\rm min}}(1 + x_+^{\rm min}) + \frac{\sqrt{C_2}}{\sqrt{D}}K_1 \left( \frac{\sqrt{D}}{T_1T_2} \right) \right) \, ds
\end{equation}
where 
\begin{equation}
 X = \frac{1}{8 m_1^2 m_2^2 K_2 \left( \frac{m_1}{T_1}\right) K_2 \left( \frac{m_2}{T_2} \right)}
\end{equation}
Note that in the limit $m_1 = m_2$, this result reduces to that found in \cite{Cheek:2021cfe}, which in turn is equal to the results in \cite{Gondolo:1990dk} if one sets $T_1 = T_2$.

\section{Dark Matter Annihilation Cross Sections}
\label{app:AppB}
The annihilation and scattering cross sections of dark matter used throughout this paper are listed below:
\begin{itemize}
  \item $s$-channel DM annihilation
  \be
  \sigma_{\bar\psi\psi\to Z' \to \bar f f}(s)=\frac{g_{\rm D}^2 g_{\rm V}^2 }{12 \pi s}\left(\frac{\left(2 m_{\rm DM}^2+s\right) \left(2 m_{ f}^2+s\right) }{ \Gamma_{Z'}^2+\left(M_{Z'}^2-s\right)^2}\right)\sqrt{\frac{s-4 m_{ f}^2}{s-4 m_{\rm DM}^2}}\,.
  \ee
  \item $t$-channel DM annihilation
  
  \bea
  \sigma_{\bar\psi\psi\to Z' Z'}(s)&=&-\frac{g_{\rm D}^4 }{8 \pi s (m_{\rm DM}^2 (s-4 M_{Z'}^2)+M_{Z'}^4)} \sqrt{\frac{s-4 M_{Z'}^2}{s-4 m_{\rm DM}^2}}\left[4 m_{\rm DM}^4+m_{\rm DM}^2 s +2 M_{Z'}^4\right.\nonumber\\
  &-&\frac{2 (m_{\rm DM}^2 \left(s-4 M_{Z'}^2\right)+M_{Z'}^4) (-8 m_{\rm DM}^4+4 m_{\rm DM}^2 (s-2 M_{Z'}^2)+4 M_{Z'}^4+s^2)
  }{\sqrt{s-4 m_{\rm DM}^2} \sqrt{s-4 M_{Z'}^2} (s-2 M_{Z'}^2)}\times\nonumber\\
  &&\left.\coth ^{-1}\left(\frac{s-2 M_{Z'}^2}{\sqrt{s-4 m_{\rm DM}^2} \sqrt{s-4 M_{Z'}^2}}\right)\right]  
  \eea
  
  \item $t$-channel DM scattering off SM particles
  \bea
  \sigma_{\psi f\to \psi f}(s)&=&\frac{g_{\rm D}^2g_{\rm V}^2}{16 \pi s}\left[2+\frac{4 s (M_{Z'}^2+s)}{m_{\rm DM}^4-2 m_{\rm DM}^2 (m_f^2+s)+(m_f^2-s)^2} \right.\times\nonumber\\
  &&\log \left(\frac{M_{Z'}^2 s}{m_{\rm DM}^4-2 m_{\rm DM}^2 (m_f^2+s)+m_f^4-2 m_f^2 s+s (M_{Z'}^2+s)}\right)\nonumber\\
  &+&\left.\frac{2 s}{M_{Z'}^2} \left(\frac{8 m_{\rm DM}^2 m_f^2+M_{Z'}^4}{m_{\rm DM}^4-2 m_{\rm DM}^2 (m_f^2+s)+m_f^4-2 m_f^2 s+s (M_{Z'}^2+s)}+2\right)\right]\nonumber\\
  \eea
\end{itemize}
In these expressions, the width of the $Z'$ is given by
\bea
\Gamma_{Z'}&=&\frac{g_{\rm D}^2 \sqrt{M_{Z'}^2-4 m_{\rm DM}^2} \left(2 m_{\rm DM}^2+M_{Z'}^2\right)}{12 \pi M_{Z'}^2}\theta (M_{Z'}^2-4 m_{\rm DM}^2)\nonumber\\
&+&\frac{g_{\rm V}^2 \sqrt{M_{Z'}^2-4 m_{\rm SM}^2} \left(2 m_{\rm SM}^2+M_{Z'}^2\right)}{12 \pi M_{Z'}^2}\theta (M_{Z'}^2-4 m_{\rm SM}^2)\,.
\eea
\end{appendix}

\bibliographystyle{JHEP}
\bibliography{Bibliography.bib}

\end{document}